\documentclass[aps,prc,floatfix,showpacs,twocolumn, preprintnumbers]{revtex4-2}

\usepackage{caption}
\usepackage{subcaption}
\usepackage{graphicx}
\usepackage{amssymb}
\usepackage{amsmath}
\usepackage{cancel}
\usepackage{placeins}
\usepackage[dvipsnames,usenames]{xcolor}
\usepackage{dcolumn}
\usepackage{bm}
\usepackage{physics}
\usepackage{hyperref}
\usepackage{float}
\usepackage{soul}

\usepackage[utf8]{inputenc}

\usepackage{bm}
\usepackage{color}
\usepackage{graphicx,epsfig}
\usepackage{amsfonts,amssymb,amsmath}
\usepackage{float}
\usepackage{xcolor}
\usepackage{color}
\usepackage{multirow}
\usepackage{subcaption}

\usepackage{dcolumn}     

\newcommand{\nc}{\newcommand}

\newcommand{\beq}{\begin{equation}}
\newcommand{\eeq}{\end{equation}}
\nc{\bfx}{{\bf x}}
\nc{\bfy}{{\bf y}}
\nc{\bfz}{{\bf z}}
\nc{\bfxh}{{\bf \hat{x}}}
\nc{\bfyh}{{\bf \hat{y}}}
\nc{\bfzh}{{\bf \hat{z}}}
\nc{\bfj}{{\bf j}}
\nc{\bfr}{{\bf r}}
\nc{\bfR}{{\bf R}}
\nc{\bfk}{{\bf k}}
\nc{\bfq}{{\bf q}}
\nc{\bfp}{{\bf p}}
\nc{\bfv}{{\bf v}}
\nc{\bfs}{{\bf s}}
\nc{\bfA}{{\bf A}}
\nc{\bfJ}{{\bf J}}
\nc{\bfsg}{{\bm \sigma}}
\nc{\bfta}{{\bm \tau}}
\nc{\bfvh}{{\bf \hat{v}}}
\nc{\bfqh}{{\bf \hat{q}}}
\nc{\low}{\delta_{\rm Low}}
\newcommand{\inner}[2]{\langle #1 | #2 \rangle}
\newcommand{\avg}[1]{\langle #1 \rangle}

\nc{\swap}{\rightleftharpoons}

\newcommand{\het}{{}^3\text{He}}
\newcommand{\tri}{{}^3\text{H}}
\newcommand{\lisix}{{}^6\text{Li}}
\newcommand{\lisev}{{}^7\text{Li}}
\newcommand{\lieight}{{}^8\text{Li}}
\newcommand{\beight}{{}^8\text{B}}
\newcommand{\besev}{{}^7\text{Be}}
\newcommand{\linin}{{}^9\text{Li}}
\newcommand{\cnin}{{}^9\text{C}}
\newcommand{\benin}{{}^9\text{Be}}
\newcommand{\bnin}{{}^9\text{B}}
\newcommand{\bten}{{}^{10}\text{B}}

\newcolumntype{d}[1]{D{.}{.}{#1}}

\newcommand{\mc}[1]{\multicolumn{1}{c}{#1}}

\def\beq{\begin{equation}}
\def\eeq{\end{equation}}
\def\beqy{\begin{eqnarray}}
\def\eeqy{\end{eqnarray}}

\begin{document}

\title{Magnetic structure of $A \le 10$ nuclei using the Norfolk nuclear models with quantum Monte Carlo methods}

\author{G. \ Chambers-Wall$^{1}$}
\email{chambers-wall@wustl.edu}
\author{A. \ Gnech$^{2,3}$}
\email{agnech@odu.edu}
\author{G. B. \ King$^{1}$}
\email{kingg@wustl.edu}
\author{S.\ Pastore$^{1,4}$}
\email{saori@wustl.edu}
\author{M.\ Piarulli$^{1,4}$}
\email{m.piarulli@wustl.edu}
\author{R.\ Schiavilla$^{2,3}$}
\email{schiavil@jlab.org}
\author{\mbox{R. B.\ Wiringa$^5$}}
\email{wiringa@anl.gov}

\affiliation{
$^1$\mbox{Department of Physics, Washington University in Saint Louis, Saint Louis, MO 63130, USA}\\
$^2$\mbox{Department  of  Physics,  Old  Dominion  University,  Norfolk,  VA  23529}\\
$^3$\mbox{Theory  Center,  Jefferson  Lab,  Newport  News,  VA  23610}\\
$^4$\mbox{McDonnell Center for the Space Sciences at Washington University in St. Louis, MO 63130, USA}\\
$^5$\mbox{Physics Division, Argonne National Laboratory, Argonne, IL 60439}\\
}

\begin{abstract}
We present Quantum Monte Carlo calculations of magnetic moments, form factors, and densities of $A\le 10$ nuclei within a chiral effective field theory approach. We use the Norfolk two- and three-body chiral potentials and their consistent electromagnetic one- and two-nucleon current operators. We find that two-body contributions to the magnetic moment can be large (up to $\sim33\%$ in $A=9$ systems). We study the model dependence of these observables and place particular emphasis on investigating their sensitivity to using different cutoffs to regulate the many-nucleon operators. Calculations of elastic magnetic form factors for $A\leq 10$ nuclei show excellent agreement with the data out to momentum transfers $q\approx 3$ fm$^{-1}$.
\end{abstract}

\maketitle

\section{Introduction}
\label{sec:intro}

In this work, we present quantum Monte Carlo (QMC) calculations, including both variational Monte Carlo (VMC) and Green's Function Monte Carlo (GFMC) methods, to study magnetic moments and form factors of $A\le 10$ nuclei. We use the precise experimental measurements of these electromagnetic observables to validate our microscopic modeling; namely, we verify the accuracy of the approach in which we treat the nucleus as a system of correlated nucleons interacting amongst themselves via two- and three-nucleon forces. Additionally, we include the coupling to external electromagnetic fields via one- and two-nucleon currents.

The specific models upon which we base this study are the Norfolk two- and three-nucleon potentials~\cite{Piarulli:2014bda,Piarulli:2016vel,Piarulli:2017dwd,Baroni:2018fdn,Piarulli:2019cqu}, derived from a $\chi$EFT that retains pions, nucleons, and $\Delta$'s as explicit degrees of freedom. The Norfolk two- and three-nucleon interactions are formulated in configuration space and we denote them as NV2+3 throughout the reminder of this work. The associated electromagnetic currents -- derived within the same $\chi$EFT framework with pions, nucleons, and $\Delta$'s -- were most recently reported in Refs.~\cite{Schiavilla:2018udt,Gnech:2022vwr}, where they were obtained up to next-to-next-to-next-to-leading order (N3LO) in the chiral expansion. The LO contribution consists of the standard single-nucleon convection and spin-magnetization terms, while sub-leading two-nucleon currents include contributions of one- and two-pion range, as well as contact terms encoding short-range dynamics~\cite{Schiavilla:2018udt}.

Previous QMC calculations of magnetic moments and transitions in $A\le 9$ systems~\cite{Pastore:2012rp} highlighted the importance of sub-leading two-nucleon currents to reach agreement with the available experimental data. The studies of Ref.~\cite{Pastore:2012rp} adopt the so-called `hybrid approach' where nuclear wave functions were determined from realistic two- and three-nucleon interactions, specifically the Argonne AV18~\cite{Wiringa:1994wb} and the Illinois-7~\cite{doi:10.1063/1.2932280}, respectively. Meanwhile, the electromagnetic currents were derived from $\chi$EFT~\cite{Pastore:2008ui,Pastore:2009is,Pastore:2011ip,Piarulli:2012bn}, thus under-utilizing the full potential of the $\chi$EFT framework to make systematic determinations of model dependencies. For instance, recent calculations of magnetic moments, based on both $\chi$EFT Hamiltonians and consistently derived $\chi$EFT electromagnetic current operators~\cite{Gnech:2022vwr,Martin:2023dhl,Pal:2023gll}, indicated a sizable sensitivity to the regulators inherent in the theoretical approach. Further, the studies in Ref.~\cite{Gnech:2022vwr,Martin:2023dhl} also showed that the N3LO contribution to the magnetic moment is enhanced relative to the size expected from naive dimensional analysis. 

By  reexamining  the  evaluation  of the magnetic structure of light nuclei using the NV2+3 model,  in  combination  with  consistent  chiral  electromagnetic currents, we aim to verify the accuracy of our $\chi$EFT approach, understand its kinematic regime of applicability, and examine the convergence of the power expansion. As in the study of Gamow-Teller $\beta$-decays reported in Ref.~\cite{King:2020wmp}, we similarly perform an in-depth analysis of two-nucleon magnetic densities. These quantities prove again to be instrumental in determining model dependencies, identifying universal short-range behaviors, and understanding the role played by inherent parameters in the models, such as low-energy constants and regulators.

Along with magnetic moments, the framework described above also allows for the study of magnetic form factors, as has already been performed for $A\le3$ systems~\cite{Gnech:2022vwr}. In particular, the work of Ref.~\cite{Gnech:2022vwr}, based on the  hyperspherical harmonics method, showed excellent agreement with data for magnetic form factors of $A\le 3$ nuclei up to momentum transfers $q\sim4$ fm$^{-1}$, which is well beyond the region of convergence for $\chi$EFT. Here, we extend these calculations to nuclei with mass number up to $A=10$ with the VMC method to further investigate this high-momentum behavior. We note that these are the first QMC calculations of form factors in $A=7-10$ systems. 

All together, the studies of electromagnetic properties in this work provide a benchmark of electroweak current models against readily available data over a wide range of kinematics. Such a wide-ranging validation is necessary, as an accurate electromagnetic current model is needed for studies of electromagnetic transitions~\cite{Wiringa:1998hr,Pastore:2012rp,Pastore:2014oda,Stroberg:2022ltv,Acharya:2023ird}, electron-nucleus scattering~\cite{Lovato:2013cua,Lovato:2015qka,Lovato:2016gkq,Pastore:2019urn,Andreoli:2021cxo,Sobczyk:2023mey,Sobczyk:2023sxh}, and radiative corrections to super-allowed beta decays~\cite{Seng:2022cnq,Cirigliano:2024rfk,Gennari:2024sbn}. These models are also needed to interpret future searches for new physics~\cite{King:2024zbv}, including precision beta-decay at low energy- and momentum-transfer~\cite{Gonzalez-Alonso:2018omy,Falkowski:2020pma,King:2022zkz,Glick-Magid:2021uwb,Brodeur:2023eul,King:2024zbv}, neutrinoless double beta decay at moderate momentum-transfers~\cite{Engel:2016xgb,Agostini:2022zub,Pastore:2017ofx,Cirigliano:2019vdj}, and long-baseline neutrino oscillation experiments that will involve both high energy- and momentum-transfer neutrino-nucleus scattering~\cite{Abe:2019,NOvA:2019cyt,Acciarri:2017,Aliaga:2014,Seo:2018,Abi:2020,Ruso:2022qes}. These analyses rely on theoretical inputs in order to disentangle signals of new physics from nuclear physics effects; thus, an accurate understanding of the underlying nuclear dynamics is necessary. Therefore, the computation of electromagnetic properties for light nuclei and detailed analysis of model dependencies are pivotal for the future of programs in both nuclear and fundamental science. 

This work is organized as follows: In Section~\ref{sec:theory}, we detail the theoretical basis for the calculation. We review the QMC computational technique in Section~\ref{sec:qmc} and report on the NV2+3 interaction models and associated electromagnetic current used in this work in Sections~\ref{sec:nv23} and~\ref{sec:current}, respectively. Section~\ref{sec:multipole} contains the multipole analysis relating QMC matrix elements to electromagnetic observables. Magnetic moments and two-body magnetic densities computed in this formalism are presented in Section~\ref{sec:mm}. This is followed, in Section~\ref{sec:ff}, by a detailed discussion of magnetic form factors that supplements the results and extends the analysis of a PRL submitted concurrently to this article~\cite{prl}. Finally, we give concluding remarks in Section~\ref{sec:conclusion}.

\section{Theory}
\label{sec:theory}

\subsection{Quantum Monte Carlo methods}
\label{sec:qmc}

Quantum Monte Carlo (QMC) approaches~\cite{Carlson:2014vla} are a suite of computational methods used to stochastically solve the Schr\"{o}dinger equation 
\begin{equation}
H \Psi(J^\pi;T,T_z)= E \Psi(J^\pi;T,T_z) \ ,
\end{equation}
where $\Psi(J^\pi;T,T_z)$ is a nuclear wave function with specific spin-parity 
$J^\pi$, isospin $T$, and charge state $T_z$.
The Hamiltonian has the form
\beqy \label{eq:hamiltonian} H = \sum_{i} K_i + {\sum_{i<j}} v_{ij} + \sum_{i<j<k}
V_{ijk} \ ,
\eeqy
where $K_i$ is the non-relativistic kinetic energy and $v_{ij}$ and $V_{ijk}$
are respectively two- and three-nucleon potentials.

The QMC approach of this work is a two-step process that begins with the variational Monte Carlo (VMC) method. In nuclear VMC calculations, the variational ansatz taken for the many-body wave function is~\cite{Pudliner:1997ck}
\beq
\ket{\Psi_T} = \mathcal{S}\prod_{i<j}\left[ 1 + U_{ij} + \sum_{i<j\neq k} \widetilde{U}^{\rm TNI}_{ijk} \right] \ket{\Psi_J}\, \label{eq:psi.t}
\eeq
where $\mathcal{S}$ is the symmetrization operator, $U_{ij}$ is a two-body correlation operator, $\widetilde{U}^{\rm TNI}_{ijk}$ is a three-body correlation operator, and $\Psi_J$ is a Jastrow-like wave function.
The antisymmetric state $\Psi_J$ encodes the long-range structure of the system and is constructed by acting on a Slater determinant that places nucleons in $s$- and $p$-shell orbitals with correlation functions that encode the appropriate cluster structure ({\it e.g.}, $^6$Li $\sim \alpha$ + $d$) of the system. 
The design of the two- and three-body correlation functions in Eq.~\ref{eq:psi.t} reflects the influence of the short-distance nuclear interaction in the medium. The correlation operators contain embedded variational parameters that one then optimizes to find a best variational state, $\Psi_V$, by minimizing the expectation value 
\beq
E_V = \frac{\mel{\Psi_V}{H}{\Psi_V}}{\inner{\Psi_V}{\Psi_V}} \geq E_0
\eeq
where $E_0$ is the true ground state energy of the system. 

The VMC result is then further improved upon by using the Green's function Monte Carlo (GFMC) method. The method leverages the fact that the real time ($t$) Schr\"{o}dinger Equation 
\beq
i\frac{\partial}{\partial t}\ket{\Psi(t)} = (H-E_T)\ket{\Psi(t)}\, ,
\eeq
may be recast as a diffusion equation in imaginary time $\tau=it$
\beq
-\frac{\partial}{\partial \tau}\ket{\Psi(\tau)} = (H-E_T)\ket{\Psi(\tau)}\, ,
\label{eq:diffusion}
\eeq
where $E_T$ is an energy offset that controls the normalization of the wave function. 
Noting that one may expand any state in a complete orthonormal basis, we can write $\Psi_V$ as a sum of the true eigenstates $\psi_i$ of the Hamiltonian 
\beq
\ket{\Psi_V} = \sum_{i=1}^{\infty} c_i\ket{\psi_i}\, .
\eeq
From Eq.~(\ref{eq:diffusion}), if the energy offset $E_T$ is taken to be the exact energy of the ground state $E_0$, it is clear that in the limit $\tau \to \infty$
\beq
\lim_{\tau\to\infty} e^{-(H-E_0)\tau}\ket{\Psi_V} \propto c_0\psi_0\, , 
\eeq
making it possible to project the true ground state out of $\Psi_V$.
In practice, the propagation is performed in $n$ small steps in imaginary time $\Delta\tau$,
\beq
\ket{\Psi(\tau)} = \left[ e^{-(H-E_0)\Delta\tau}\right]^n\ket{\Psi_V}\, ,
\eeq
until spurious contamination is removed from the wave function and convergence is reached.

In this work, we will study matrix elements of operators $\mathcal{O}$ acting on the VMC wave function. To obtain expectation values in GFMC, we compute the so-called ``mixed-estimate,"
\beq
\avg{\mathcal{O}(\tau)} \approx 2\frac{\mel{\Psi_V}{\mathcal{O}}{\Psi(\tau)}}{\inner{\Psi_V}{\Psi(\tau)}}
- \frac{\mel{\Psi_V}{\mathcal{O}}{\Psi_V}}{\inner{\Psi_V}{\Psi_V}}\, ,
\eeq
 derived assuming the GFMC wave function at $\tau$ is a small improvement over the VMC state; {\it i.e.}, $\Psi(\tau)=\Psi_V+\delta\Psi$. 
Since the variational wave functions are typically good descriptions of light nuclei, this a valid approximation. In principle, it is possible to calculate a matrix element between two propagated states in GFMC; however, this requires performing separate imaginary time propagations for each matrix element that one would like to compute~\cite{Pervin:2007sc}, which makes an order-by-order analysis of the magnetic moments and form factors quite costly. Thus, the mixed estimate is a necessary approximation to systematically analyze the convergence of $\chi$EFT predictions for the magnetic structure of nuclei over a range of momenta.

\subsection{The Norfolk interaction model}
\label{sec:nv23}

We base our calculations on electromagnetic many-body currents~\cite{Pastore:2008ui,Pastore:2009is,Schiavilla:2018udt,Gnech:2022vwr} derived consistently with the Norfolk two- and three-body (NV2+3) nuclear interactions~\cite{Piarulli:2014bda,Piarulli:2016vel,Piarulli:2017dwd,Baroni:2018fdn,Piarulli:2019cqu}. The Norfolk two-body potential (NV2) is a fully local interaction -- hence suitable to be utilized within QMC methods -- and includes terms up to N3LO in the $\chi$EFT expansion. To this order, the NV2 consists of  one- (OPE) and two-pion-exchange (TPE) contributions, as well as contact interactions with strengths specified by unknown low-energy constants (LECs). Non-localities emerging from the contact terms entering at N3LO have been eliminated by i) leveraging Fierz transformations to develop a minimally non-local interaction to order N3LO~\cite{Piarulli:2014bda}; and ii) constraining the LECs associated with the leftover N3LO non-local terms to vanish~\cite{Piarulli:2016vel}. This choice is motivated by the consideration that the LECs associated with the non-local terms are of natural size, as opposed to those associated with the local isovector and isotensor operators required to  obtain good agreement with nucleon-nucleon ($N\!N$) scattering data.

\begin{figure}
\begin{center}
    \includegraphics[width=1.2in]{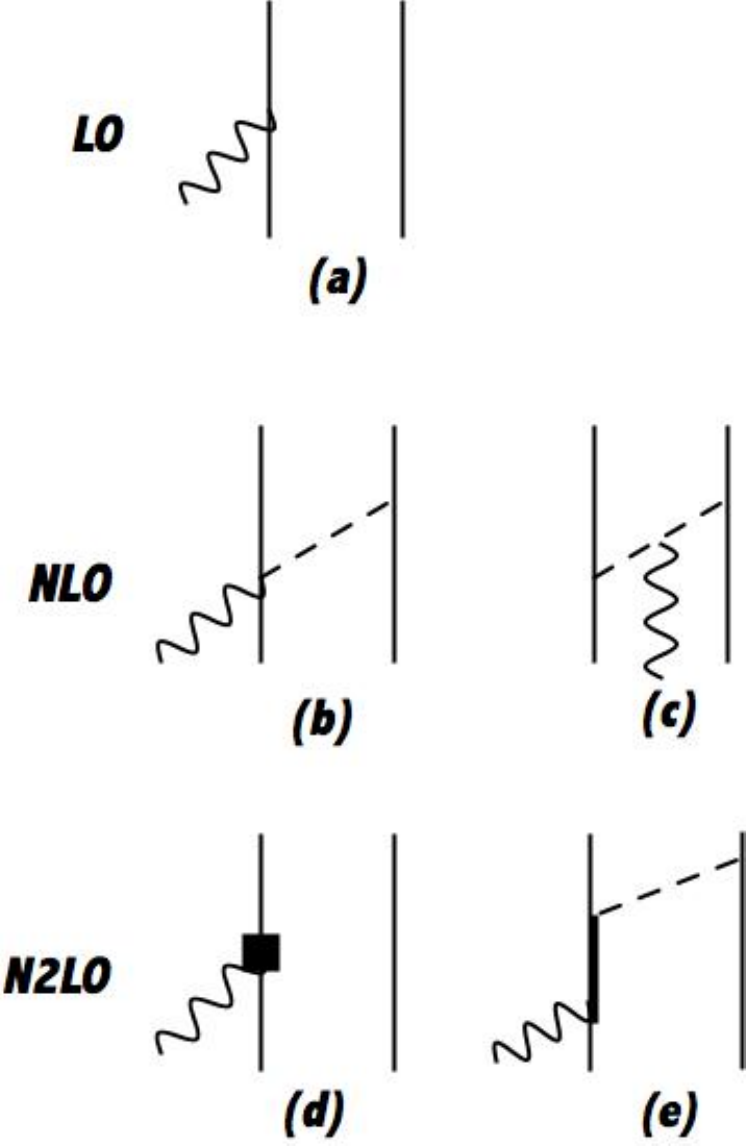}
\end{center}
\caption{Diagrams illustrating the contributions to the electromagnetic current up to N2LO. Nucleons,  $\Delta$-isobars,  pions,  and  external  fields  are  denoted  by  solid,  thick-solid,  dashed,  and  wavy  lines,  respectively.   The square in panel (d) represents relativistic corrections to the LO current. Figure from Ref.~\cite{Schiavilla:2018udt}.}
\label{fig:currn2lo}
\end{figure}

Schematically, the NV2 interaction can be broken down into three parts as
\begin{equation}
v_{12} = v_{12}^L + v_{12}^S + v_{12}^{\rm EM}\, ,
\end{equation}
where $v_{12}^L$ is the long-range term consisting of OPE and TPE contributions to order N2LO, $v_{12}^S$ is the short-range term consisting of contact 
interactions, and $v^{\rm EM}_{12}$ is an electromagnetic contribution. The electromagnetic term retains second-order Coulomb, Darwin-Foldy, vacuum polarization, and magnetic moment interaction terms~\cite{Wiringa:1994wb}. 

The radial functions in $v_{12}^L$ have ${\sim} 1/r^n$ (with $n\le6$) singularities at the origin that are regulated with the function 
\begin{equation}
C_{R_L}(r) =  1 - \frac{1}{(r/R_L)^6e^{2(r-R_L)/R_L} + 1 },
\end{equation}
where $R_L$ is the long-range cutoff. The short-range regulator comes from a Gaussian smearing of the $\delta$-functions in coordinate space, expressed as
\begin{equation}
C_{R_S}(r) = \frac{1}{\pi^{3/2}R_S^3}e^{-(r/R_S)^2}\, ,
\end{equation}
where $R_S$ is the Gaussian parameter governing the width of the smearing. Consequently, two regulator parameters need to be specified in the model. 

Embedded within $v_{12}^S$ are 26 unknown LECs that parameterize the strengths of the contact terms, adjusted to fit $N\!N$ scattering data~\cite{Perez:2013jpa,Perez:2013oba,Perez:2014yla}. There are four classes of models for the NV2. Model classes denoted by I (II) are fit to $N\!N$ scattering data up to 125 (200) MeV with a $\chi^2/{\rm datum}\approx 1.1~(1.4)$. For model classes denoted by a (b), the regulators used in the potential are $[R_L,R_S]=[1.2~{\rm fm},0.8~{\rm fm}]$ ($[R_L,R_S]=[1.0~{\rm fm},0.7~{\rm fm}]$). 

In addition to the NV2, the Norfolk models incorporate a three-body force (NV3)~\cite{vanKolck:1994yi,Epelbaum:2002vt} that requires the determination of two additional LECs; namely,  $c_D$ and $c_E$. These are fit either to the triton ground state energy and $nd$ doublet scattering length~\cite{Piarulli:2017dwd} or to the triton ground state energy and Gamow-Teller matrix element for $\beta$-decay~\cite{Baroni:2018fdn}. The latter fitting procedure corresponds to the model classes denoted with a star, which are the ones used in this work.

\subsection{The electromagnetic current operator}
\label{sec:current}

\begin{figure}
\begin{center}
    \includegraphics[width=2.2in]{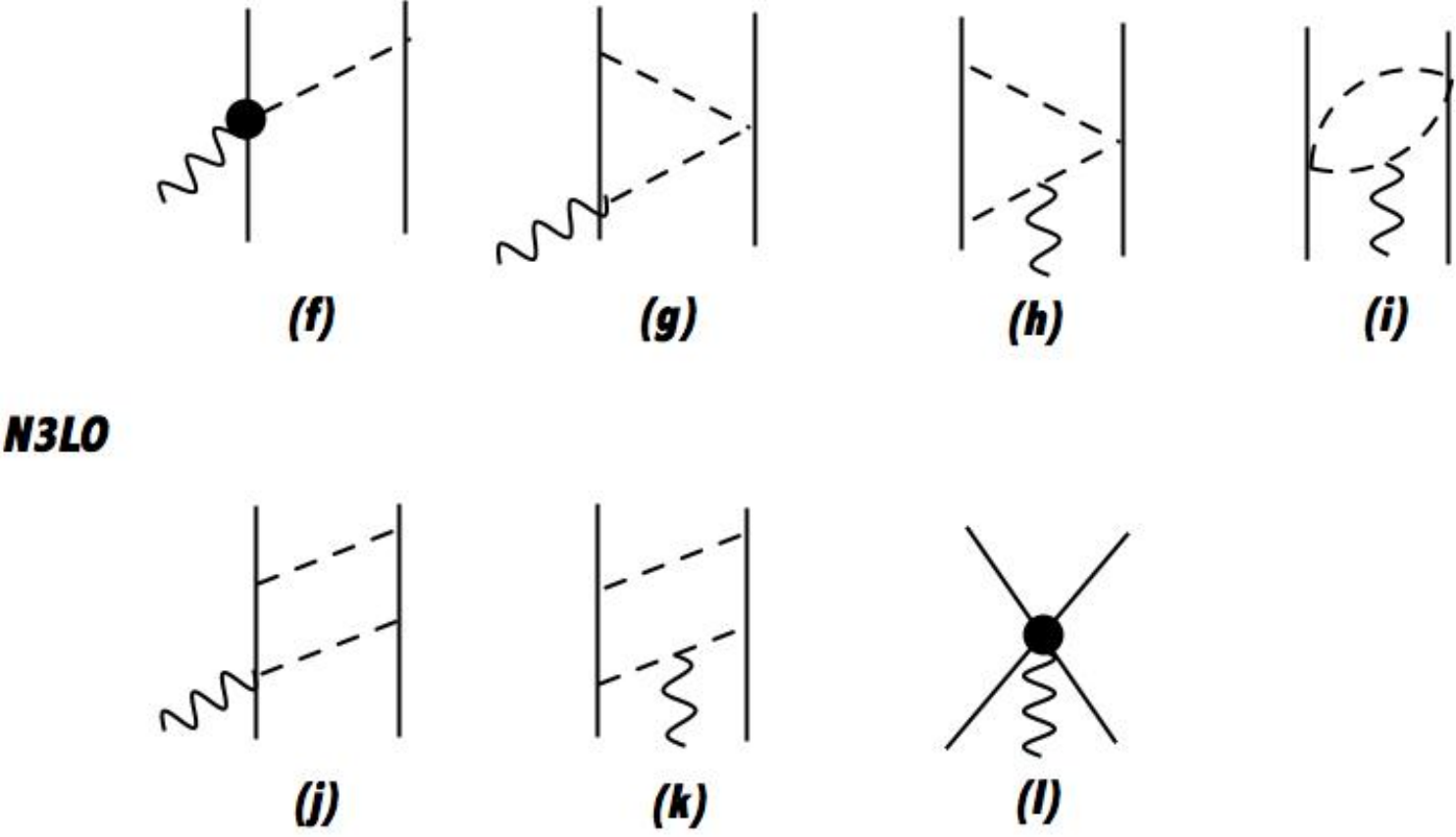}
\end{center}
\caption{Diagrams  illustrating  the  contributions  to  the  electromagnetic  current  at  N3LO adopted in this work.  Nucleons,  pions,  and  external  fields  are  denoted  by  solid,  dashed,  and  wavy  lines, respectively.  The solid circle in panel (f) is associated with a sub-leading $\gamma\pi N$ interaction. Figure from Ref.~\cite{Schiavilla:2018udt}. }
\label{fig:currn3lo}
\end{figure}

The electromagnetic current operator, ${\bf j}$, is expressed as an expansion in many-body terms as
\begin{eqnarray}
{\bf j} &=&  \sum_i {\bf j}_i({\bf q}) + \sum_{i<j} {\bf j}_{ij}({\bf q}) +\dots  \ ,
\end{eqnarray}
where ${\bf q}$ is the momentum transferred to the nucleus. In this work we include up to two-nucleon contributions. 

Electromagnetic currents have been extensively studied within several implementations of $\chi$EFT, including covariant perturbation theory~\cite{Park:1995pn}, the unitary transformation method~\cite{Kolling:2009iq,Kolling:2011mt,Krebs:2016rqz,Krebs:2019aka}, and time-ordered perturbation theory~\cite{Pastore:2008ui,Pastore:2009is,Pastore:2011ip,Piarulli:2012bn,Schiavilla:2018udt,Gnech:2022vwr}.
In this work, we adopt the operators developed using time-ordered perturbation theory~\cite{Pastore:2008ui,Pastore:2009is,Pastore:2011ip,Piarulli:2012bn,Schiavilla:2018udt,Gnech:2022vwr} whose expressions in momentum space are summarized in Ref.~\cite{Gnech:2022vwr}. For ease of reading, we detail the corresponding coordinate space expressions in Appendix~\ref{app:currr}.  

 Figs.~\ref{fig:currn2lo} and~\ref{fig:currn3lo} display schematic representations of the electromagnetic current operator. One-body operators appear at LO and N2LO. These are generated by the disconnected diagrams displayed in panels (a) and (d) of Fig.~\ref{fig:currn2lo}, respectively. The LO one-body operator consists of the convection and the spin-magnetization currents associated with an individual nucleon. This operator is derived from the non-relativistic reduction of the covariant single-nucleon current; that is to say, by expanding the operator in powers of $p_i/m_N$, where $p_i$ and $m_N$ are the momentum and mass of a nucleon, respectively. Its expression in coordinate space is given in Eq.~(\ref{eq:jlo}) and is denoted with `LO' in the tables. The one-body operator at N2LO  accounts for relativistic corrections to the LO one-body operator, and its expression can be found in Eq.~(\ref{eq:j1rcc}).

Two-body operators first appear at NLO in the expansion. At this order, they consist of a purely isovector (IV) contribution generated by the seagull and pion-flight diagrams illustrated in panels (b) and (c) of Fig.~\ref{fig:currn2lo}. The NLO two-body current is denoted with `NLO' in the tables, and its expression is provided in Eq~(\ref{eq:nlor2}). 

At N2LO, there is the two-body OPE current involving the excitation of an intermediate $\Delta$-isobar shown in panel (e) of Fig.~\ref{fig:currn2lo}. In the tables, we report the cumulative contribution at N2LO, that is the sum of panels (d) and (e), and denote it with `N2LO'. When discussing the two-body magnetic densities, we analyze the N2LO OPE current involving the intermediate $\Delta$-isobar separately, and denote its contribution with `N2LO($\Delta$)'. This OPE term is purely IV and transverse, thus unconstrained by current conservation with the NV2 interaction. 

The contributions at N3LO are shown in Fig.~\ref{fig:currn3lo}. Panel (a) displays the OPE current involving a sub-leading term in the $\gamma\pi N$ chiral Lagrangian that generates both an isoscalar (IS) and an IV current. These N3LO contributions of one-pion range are cumulatively denoted with `N3LO(OPE)' in the tables. When discussing the two-body magnetic densities, we analyze separately the effect of its individual components. Specifically, the IS component---given in Eq.~(\ref{eq:n3lo.is})---is proportional to the unknown LEC $d_9^\prime$, and is denoted with `N3LO(OPE) IS' in the figures. The IV component involves two unknown LECs, namely, $d_8^\prime$ and $d_{21}^\prime$. The term associated with $d_8^\prime$ is given in Eq.~(\ref{eq:n3lo.iv.d8}), and is labeled with `N3LO(OPE) IV $d_8^\prime$'. While the term involving $d_{21}^\prime$ is given in Eq.~(\ref{eq:n3lo.iv.d21}), and is labeled with `N3LO(OPE) IV $d_{21}^\prime$'. 

Currents of two-pion range (without $\Delta$-isobars) are shown in panels (g)--(k) of Fig.~\ref{fig:currn3lo} and are denoted with `N3LO(LOOP)' in the tables and in the figures.

The last contribution at N3LO is represented by the contact term in panel (l) of Fig.~\ref{fig:currn3lo}. We distinguish between two kinds of contact contributions, namely the minimal (MIN) and non-minimal (NM) contact currents. The former is linked to the NV2 contact potential at NLO via current conservation; therefore it involves the same LECs entering the NV2. This contribution is denoted with `N3LO(MIN)' in the tables. The N3LO(MIN) term is broken down into its IS and IV components---denoted as `N3LO(MIN) IS' and `N3LO(MIN) IV'---given in Eqs.~(\ref{eq:jmin.is}) and~(\ref{eq:jmin.iv}), respectively. 

Non-minimal contact currents are unconstrained by current conservation and are denoted with `N3LO(NM)' in the tables. The non-minimal contact currents are proportional to two unknown LECs, $C_{15}^\prime$ and $C_{16}^\prime$. The former enters the IS current given in Eq.~(\ref{eq:jnm.is}) that generates a contribution labeled as `N3LO(NM) IS'. The latter enters the IV current given in Eq.~(\ref{eq:jnm.iv}) that gives rise to a contribution labeled as  `N3LO(NM) IV'.

The operators described above involve five unknown LECs, $d_8^\prime$, $d_9^\prime$, $d_{21}^\prime$, $C_{15}^\prime$ and $C_{16}^\prime$ for which we adopt the values obtained from the analysis of Ref.~\cite{Gnech:2022vwr}. In particular, these LECs are constrained to reproduce the magnetic moments of the deuteron and the trinucleon systems, as well as backward-angle deuteron photo-disintegration data. This fitting strategy, denoted as ``fit A" in Ref.~\cite{Gnech:2022vwr}, provides excellent agreement with the data, even at large values of momentum transfer, when used to compute magnetic form factors in the hyperspherical harmonics approach. For convenience, we report the values of these LECs in Appendix~\ref{app:currr}.

\subsection{Magnetic form factors and moments}
\label{sec:multipole}

The magnetic form factors are extracted from elastic electron scattering differential cross section data (see Ref.~\cite{Donnelly1984} for a review). For this work we use the definition given in Ref.~\cite{Donnelly1984}, {\it i.e.}, 
\begin{equation}
\begin{aligned}
F_M^2(q)= \frac{1}{2 J+1} \sum_{L=1}^{\infty}\left|\left\langle J ||M_L(q)|| J\right\rangle\right|^2\,,\label{eq:fm}
\end{aligned}
\end{equation}
where we express the magnetic form factor as a function of  the reduced matrix elements (RMEs) of the magnetic  multipole operators ($M_L$), and $J$ is the total angular momentum of the nucleus. 

The calculation of the magnetic multipole expansion is carried out using standard techniques~\cite{Walecka1995} implemented to expand the current operator $\bfj(\bfq)$. 
As the reference frame, we adopt the one where $\hat{\bm{z}}$ is the spin-quantization
axis of the nucleus, and the direction of $\hat{\bfq}$ is defined by the angles $\theta$ and $\phi$, the polar and azimuthal angles with respect to this axis, respectively. In what follows, we define $q=|\bfq|$. The current, in terms of the magnetic and electric multipoles, $M_{L M_L}$ and $E_{L M_L}$,  reads
\begin{equation}
\begin{aligned}
\mathrm{j}_{\lambda}(\mathbf{q})= & \int d \mathbf{x}\, e^{i \mathbf{q} \cdot \mathbf{x}} \hat{\mathbf{e}}_{\lambda} \cdot \mathbf{j}(\mathbf{x}) \\
= & -\sum_{L \geq 1,M_L} \sqrt{2 \pi(2 L+1)} i^L D_{M_L, \lambda}^L(-\phi,-\theta, \phi)\\
&\times  \left[\lambda M_{L M_L}(q)+E_{L M_L}(q)\right], \label{eq:jmultipole}
\end{aligned}
\end{equation}
where $\lambda=\pm1$, and $D_{M_L, \lambda}^L$ are the Wigner rotation matrices~\cite{Edmonds1957}. The unit vector $\hat{\mathbf{e}}_{q \lambda}$ indicates
the linear combination
\begin{equation}
  \hat{\mathbf{e}}_{\lambda=\pm1}=\mp\frac{\hat{\mathbf{e}}_{1}\pm\hat{\mathbf{e}}_{2}}{\sqrt{2}}\,,
\end{equation}
where $\hat{\mathbf{e}}_{3}=\hat{\bfq}$, $\hat{\mathbf{e}}_{2}=\hat{\bm{z}}\times\hat{\bfq}$,
and $\hat{\mathbf{e}}_{1}=\hat{\mathbf{e}}_{2}\times\hat{\mathbf{e}}_{3}$. 
Note that, for elastic scattering, the electric multipole operators ($E_L$) vanish because of time reversal
invariance~\cite{Forest1966}, and the sum over $L$ runs only on odd values because of parity conservation.

As observed in Ref.~\cite{Carlson2015}, in VMC and GFMC calculations, it
is more efficient to compute the matrix elements  of $\bfj(\bfq)$ between states of a specific spin configuration $M_J$ (usually $M_J=J$) and then adjust the direction of $\hat{\bfq}$ to isolate the single RME contributions. In our case, it is very convenient to select $\bfq$ to be in the $x-z$ plane. This  makes $\phi=0$, and  $\hat{\mathbf{e}}_2=\hat{\bm y}$. Leveraging this selection and making use of the Wigner-Eckart theorem in Eq.~(\ref{eq:jmultipole}), the  matrix element reduces to~\cite{Carlson2015}
\begin{equation}
\begin{aligned}\label{eq:mej}
 \left\langle J J \left|\mathrm{j}_{y}(\mathbf{q})\right| J J\right\rangle
= &\sqrt{4 \pi}\sum_{L \geq 1} i^{L+1}  \frac{\langle J J, J -\!J| L 0 \rangle}{\sqrt{L(L+1)}} \\
&\times P^1_L(\cos \theta)\left\langle J|| M_L(q) \| J\right\rangle\,.
\end{aligned}
\end{equation}

The various multipole contribution to the magnetic form factor for the nuclei  considered in this work are summarized in Table~\ref{tab:tab_multi}. The number of independent matrix elements with different directions of $\bfq$ ({\it i.e.}, of different $\theta$'s) that have to be computed is equal to the number of multipoles allowed. Clearly, any independent choice of the directions are equivalent; in Appendix~\ref{sec:app1} we present the explicit formulas obtained for our specific choices.
\begin{table}[]
    \centering
    \begin{tabular}{ccc}
        \hline
        \hline
         $J^\pi$ & Nuclei & Multipoles \\
            \hline
         $1/2^+$ & $\het$,$\tri$ & $M_1$\\
         $1^+$ & $\lisix$ & $M_1$\\
         $3/2^-$ & $\lisev$, $\besev$ & $M_1$, $M_3$\\
         $2^+$ & $\lieight$, $\beight$ & $M_1$, $M_3$\\
         $3/2^-$ & $\linin$, $\cnin$, $\benin$, $\bnin$ & $M_1$, $M_3$\\
         $3^+$ & $\bten$& $M_1$, $M_3$, $M_5$\\
         \hline
         \hline
    \end{tabular}
    \caption{Multipole contributions to the magnetic form factors for the nuclei considered in this work.}
    \label{tab:tab_multi}
\end{table}

From the small $q$ behaviour of the magnetic form factor it is possible to extract the magnetic (dipole) moment of the nucleus. Starting from the definition of the magnetic moment, it is possible to show that~\cite{Carlson2015}
\begin{equation}
\left\langle J|| M_1(q)|| J\right\rangle \simeq \frac{i}{\sqrt{2 \pi} \langle J J,J -\!J |1 0\rangle} \frac{q}{2 m_N} \mu \,.
\end{equation}
Therefore, considering that the higher-order multipoles have a negligible contribution to the magnetic moment at small $q$ we can select the matrix element computed for $\theta=\pi/2$ (any choice is equivalent) and so the magnetic moment becomes 
\beq
\mu = \lim_{q\to 0} -i \frac{2m_N}{q}\mel{JJ}{\mathrm{j}_y(q \hat{\bm x})}{JJ}\,.
\label{eq:mm}
\eeq
To extract $\mu$, we evaluate $\mel{JJ}{\mathrm{j}_y(q\bfxh)}{JJ}$ for $q \in [0.00,0.25]$ fm$^{-1}$ and fit the results to a polynomial, similarly to what was done for the multipole analysis of the $^6$He $\beta$-decay spectrum in Ref.~\cite{King:2022zkz}. The linear coefficient is then used to determine the quantity of interest. 

\section{Results and Discussion}
\label{sec:results}

In this section, we summarize our calculations. Specifically, we report the results for
i) magnetic moments; ii) two-body magnetic densities; and iii) magnetic form factors of light nuclei.

\subsection{Magnetic moments}
\label{sec:mm}

\begin{table*}[tbh]
\begin{center}
\begin{tabular} {l l *{9}{d{3.3}}}\hline \hline 
&Model &\mc{LO} &\mc{NLO} &\mc{N2LO} &\mc{N3LO(MIN)} &\mc{N3LO(NM)} &\mc{N3LO(OPE)} &\mc{N3LO(LOOP)} &\mc{TOT} &\mc{TOT-LO} \\ \hline
$^3{\rm H}(\frac{1}{2}^+;\frac{1}{2})$ &Ia$^{\star}$ &2.588 &0.196 &0.036 &0.042 &0.089 &-0.005 &0.027 &2.973 &0.385\\
&IIa$^{\star}$ &2.588 &0.197 &0.035 &0.038 &0.125 &-0.038 & 0.027 & 2.972 &0.384 \\
&Ib$^{\star}$ &2.589 &0.224 &0.061 &0.042 &0.076 &-0.030 &0.021 &2.972 &0.394\\
&IIb$^{\star}$ &2.592 &0.226 &0.059 &0.033 &0.072 &-0.014 &0.020 &2.986 &0.394 \\
&Exp & & & & & & & &2.979 &\\
\hline
$^3{\rm He}(\frac{1}{2}^+;\frac{1}{2})$ &Ia$^{\star}$ &-1.766 &-0.193 &-0.042 &0.030 &-0.117 &-0.002 &-0.027 &-2.116 &-0.350\\
&IIa$^{\star}$ &-1.767 &-0.194 &-0.043 &0.029 &-0.146 &0.032 & -0.026 & -2.116 &-0.358\\
&Ib$^{\star}$ &-1.770 &-0.220 &-0.066 &0.033 &-0.124 &0.045 &-0.026 &-2.124 &-0.358\\
&IIb$^{\star}$ & -1.769 &-0.222 &-0.066 &0.026 &-0.102 &0.026 &-0.019 &-2.127 &-0.349 \\
&Exp & & & & & & & &-2.127 &\\
\hline
$^6{\rm Li}(1^+;0)$ &Ia$^{\star}$ &0.826 & 0.000 &-0.010 &0.046 &-0.019 &-0.003 &0.000 &0.840 &0.014\\
 &IIa$^{\star}$ &0.823 & 0.000 &-0.010 &0.043 &-0.016 &-0.002 &0.000 &0.838 &0.016\\
&Ib$^{\star}$ &0.820 &0.000 &-0.010 &0.037 &-0.015 &0.001 &0.000 &0.833 &0.015 \\
&IIb$^{\star}$ &0.825 &0.000 &-0.010 &0.034 &-0.019 &0.008 &0.000 &0.838 &0.013 \\
&Exp & & & & & & & &0.822 &\\
\hline
$^7{\rm Li}(\frac{3}{2}^-;\frac{1}{2})$ &Ia$^{\star}$ &2.923 &0.184 &0.021 &0.046 &0.103 &-0.007 &0.033 &3.301 &0.379\\
&IIa$^{\star}$ &2.895 &0.187 &0.021 &0.043 &0.141 &-0.038 &0.032 &3.280 &0.385\\
&Ib$^{\star}$ &2.905 &0.224 &0.052 &0.046 &0.092 &-0.032 &0.029 &3.317 &0.413\\
&IIb$^{\star}$ &2.928 &0.220 &0.043 &0.037 &0.083 &-0.014 &0.026 &3.322 &0.394\\
&Exp & & & & & & & &3.256 &\\
\hline
$^7{\rm Be}(\frac{3}{2}^-;\frac{1}{2})$ &Ia$^{\star}$ &-1.094 &-0.183 &-0.033 &0.030 &-0.132 &0.002 &-0.033 &-1.443 &-0.349\\
&IIa$^{\star}$ &-1.070 &-0.187 &-0.035 &0.032 &-0.169 &0.032 &-0.031 &-1.465 &-0.359\\
&Ib$^{\star}$ &-1.083 &-0.222 &-0.065 &0.037 &-0.148 &0.049 &-0.029 &-1.461 &-0.359\\
&IIb$^{\star}$ &-1.098 &-0.221 &-0.060 &0.030 &-0.119 &0.027 &-0.025 &-1.464 &-0.367\\
&Exp & & & & & & & &-1.465 &\\
 \hline \hline
\end{tabular}
\end{center}
\caption{VMC results in units of $\mu_N$ for magnetic moments of nuclei with mass number $A=3-7$, computed with the NV2+3 interaction and consistent electromagnetic current broken down order-by-order in the chiral expansion. The column denoted by TOT is the sum of all contributions to the magnetic moment. The columns denoted with LO and TOT-LO indicate the contribution to the magnetic moment from the electromagnetic current at LO and the cumulative contribution from sub-leading electromagnetic currents, respectively. Statistical uncertainties from the Monte Carlo integration, which are $\lesssim 2\%$ for all contributions listed, have been omitted. The experimental data are from the evaluations of Refs.~\cite{Purcell:2010hka,Tilley:2002vg,Tilley:2004zz,Borremans:2005ar}.}
\label{tab:vmc.mm.light37}
\end{table*}

\begin{table*}[tbh]
\begin{center}
\begin{tabular} {l l *{9}{d{3.3}}}\hline \hline 
&Model &\mc{LO} &\mc{NLO} &\mc{N2LO} &\mc{N3LO(MIN)} &\mc{N3LO(NM)} &\mc{N3LO(OPE)} &\mc{N3LO(LOOP)} &\mc{TOT} &\mc{TOT-LO} \\ \hline
$^8{\rm Li}(2^+;1)$ &Ia$^{\star}$ &1.332 &0.183 &0.016 &0.064 &0.080 &-0.046 &0.031 &1.660 &0.328\\
&IIb$^{\star}$ &1.325 &0.208 &0.035 &0.049 &0.057 &-0.018 &0.022 &1.678 &0.353\\
&Exp & & & & & & & &1.654 &\\
\hline
$^8{\rm B}(2^+;1)$ &Ia$^{\star}$ &1.330 &-0.176 &-0.047 &0.054 &-0.127 &0.026 &-0.032 &1.028 &-0.302 \\
&IIb$^{\star}$ &1.274 &-0.206 &-0.065 &0.040 &-0.108 &0.031 &-0.023 &0.943 &-0.331\\
&Exp & & & & & & & &1.036 &\\
\hline
$^9{\rm Li}(\frac{3}{2}^-;\frac{3}{2})$ &Ia$^{\star}$ &2.933 &0.271 &0.031 &0.094 &0.113 &0.026 &0.027 &3.494& 0.561\\
&IIb$^{\star}$ &3.058 &0.310 &0.055 &0.068 &0.084 &-0.005 &0.006 &3.574 &0.517\\
&Exp & & & & & & & &3.437 &\\
\hline
$^9{\rm Be}(\frac{3}{2}^-;\frac{1}{2})$ &Ia$^{\star}$ &-0.964 &-0.041 &0.011 &0.010 &-0.053 &0.018 &-0.010 &-1.030 &-0.066 \\
&IIb$^{\star}$ &-1.049 &-0.049 &0.004 &0.011 &-0.044 &0.011 &-0.007 &-1.123 &-0.074\\
&Exp & & & & & & & &-1.178 &\\
\hline
$^9{\rm B}(\frac{3}{2}^-;\frac{1}{2})^{\dagger}$ &Ia$^{\star}$ &2.690 &0.040 &-0.024 &0.024 &0.039 &-0.015 &0.010 &2.763 &0.073  \\
&IIb$^{\star}$ &2.821 &0.049 &-0.017 &0.015 &0.029 &-0.005 &0.007 &2.900 &0.079\\
&Exp & & & & & & & &\mc{--} &\\
\hline
$^9{\rm C}(\frac{3}{2}^-;\frac{3}{2})^{\dagger}$ &Ia$^{\star}$ &-1.070 &-0.264 &-0.047 &0.002 &-0.150 &-0.033 &-0.028 &-1.590 &-0.520\\
&IIb$^{\star}$ &-0.983 &-0.329 &-0.079 &0.003 &-0.131 &0.020 &-0.007 &-1.505 &-0.523\\
&Exp & & & & & & & &-1.391 &\\
\hline
$^{10}{\rm B}(3^+;0)$ &Ia$^{\star}$ &1.816 &0.000 &-0.019 &0.069 &-0.028 &0.000 &0.000 &1.839 &0.023\\
&IIb$^{\star}$ &1.819 &0.000 &-0.019 &0.048 &-0.027 &0.012 &0.000 &1.833 &0.014\\
&Exp & & & & & & & &1.801 &\\ \hline \hline
\end{tabular}
\end{center}
\caption{Same as Tab.~\ref{tab:vmc.mm.light810} but for nuclei with mass number $A=8-10$.}
\label{tab:vmc.mm.light810}
\end{table*}

\begin{figure}[tbh]
\begin{center}
    \includegraphics[width=0.45\textwidth]{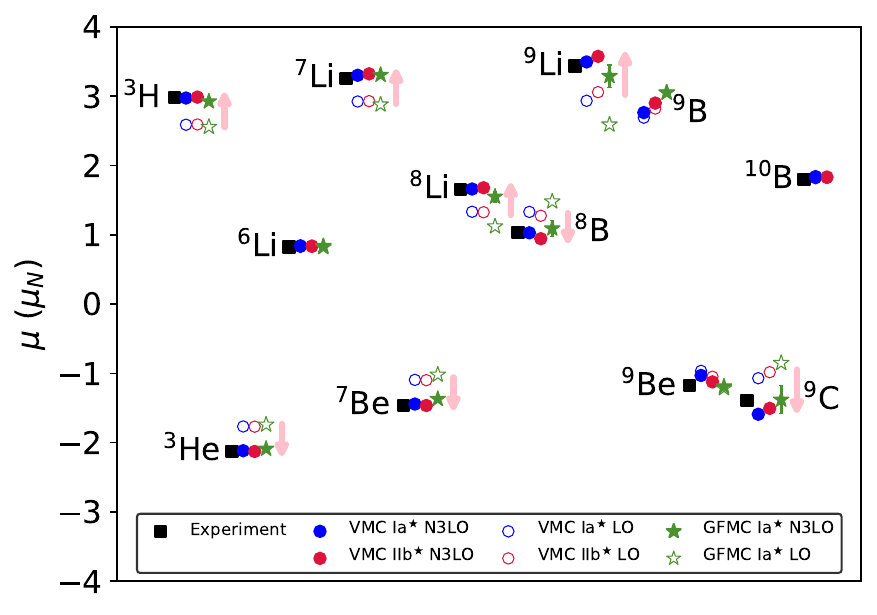}
\end{center}
\caption{VMC magnetic moments calculated with the NV2+3-Ia$^{\star}$ (blue circles) and NV2+3-IIb$^{\star}$ (red circles) compared with experiment (black dots). GFMC results based on NV2+3-Ia$^{\star}$ are given by the green stars. Empty symbols denote the calculation up to LO in the current, while filled symbols include all contributions through N3LO.}\label{fig:exp.comp}
\end{figure}

We perform QMC calculations of magnetic moments for several light nuclei with the NV2+3 interactions and associated electromagnetic currents. For nuclei with mass number $A=3-7$, we present VMC results based on four NV2+3 model classes; namely, models Ia$^{\star}$, IIa$^{\star}$, Ib$^{\star}$, and IIb$^{\star}$. In the cases of nuclei with mass number $A=8-10$, we base our studies on two models-- models Ia$^{\star}$ and IIb$^{\star}$. This approach allows us to determine the impacts resulting from variations in the cutoff parameters and energy range utilized for fitting the NV2 potential on the calculated magnetic moments. The VMC results, along with a breakdown into the various contributions from the one- and two-body chiral currents described in Sec.~\ref{sec:current}, are summarized in Fig.~\ref{fig:exp.comp} and Tables~\ref{tab:vmc.mm.light37} and~\ref{tab:vmc.mm.light810} for systems with $A=3-7$, and $A=8-10$, respectively. Fig.~\ref{fig:exp.comp} displays our results for VMC and GFMC magnetic moments with circles and stars, respectively. Calculations based on the one-body operator alone are indicated with empty symbols, while those that include two-body currents up to N3LO are indicated with filled symbols. Pink arrows indicate the major shifts induced by the two-body currents, which are positive (negative) for neutron-rich (proton-rich) nuclei. Finally, for comparison, we plot the experimental data with black squares. 

\subsubsection{VMC magnetic moments of $A=3-7$ nuclei} 

The computation of the $^3$H and $^3$He magnetic moments represents a benchmark with the exact hyperspherical harmonics calculation of Ref.~\cite{Gnech:2022vwr}. The VMC calculations are in excellent agreement with the previous results, and thus we do not summarize again the findings contained therein. The salient features of these systems are a strong two-body component of $\sim13\%~(16\%)$ to the overall $^3{\rm H}~(^3{\rm He})$ magnetic moments, as well as a $\lesssim 0.5\%$ model dependence due to the inclusion of trinucleon magnetic moments in the fit to determine the unknown electromagnetic currents LECs.

\begin{table*}[tbh]
\begin{center}
\begin{tabular} {c c *{4}{d{3.6}}}\hline \hline 
$(J^\pi;T)$& Method&\mc{LO} &\mc{TOT-LO} &\mc{TOT} &\mc{Expt.} \\ \hline
$^3{\rm H}(\frac{1}{2}^+;\frac{1}{2})$ &VMC &2.582 &0.383 &2.970 &2.9790\\
&GFMC &2.556 &0.368 &2.924 &\\ 
$^3{\rm He}(\frac{1}{2}^+;\frac{1}{2})$ &VMC &-1.766 &-0.350 &-2.116 &-2.1275\\
&GFMC &-1.740 & -0.349 &-2.089 &\\
$^6{\rm Li}(1^+;0)$ &VMC &0.826 &0.014 &0.840 &0.8221\\
&GFMC &0.821 &0.016 &0.837 &\\
$^7{\rm Li}(\frac{3}{2}^-;\frac{1}{2})$ &VMC &2.923 &0.379 &3.301 &3.2564\\
&GFMC &2.879 &0.429 &3.307 &\\
$^7{\rm Be}(\frac{3}{2}^-;\frac{1}{2})$ &VMC &-1.094 &-0.349 &-1.443 &-1.465 \\
&GFMC &-1.019(3) &-0.35(2) &-1.37(2) \\
$^8{\rm Li}(2^+;1)$ &VMC &1.332 &0.328 &1.660 &1.654 \\
&GFMC &1.12(3) &0.43(8) &1.55(8) &\\
$^8{\rm B}(2^+;1)$ &VMC &1.479 &-0.302 &1.028 &1.036\\
&GFMC &1.48(2) &-0.39(8) &1.09(11) &\\
$^9{\rm Li}(\frac{3}{2}^-;\frac{3}{2})$ &VMC &2.9331 &0.5609 &3.4939 &3.4391\\
&GFMC &2.59(2) &0.70(16) &3.29(16) & \\
$^9{\rm Be}(\frac{3}{2}^-;\frac{1}{2})$ &VMC &-0.964 & -0.066 & -1.030 &-1.178\\
&GFMC &-1.214(5) &0.02(3) &-1.19(3) & \\
$^9{\rm B}(\frac{3}{2}^-;\frac{1}{2})$ &VMC &2.690 & 0.073 & 2.763 &\mc{--} \\
&GFMC &3.054(7) &-0.01(3) &3.05(3)\\
$^9{\rm C}(\frac{3}{2}^-;\frac{3}{2})$ &VMC &-1.070 & -0.520 &-1.590 &-1.391\\
&GFMC &-0.85(3) &-0.53(19) & -1.38(20) \\ \hline \hline
\end{tabular}
\end{center}
\caption{GFMC calculations of magnetic moments of $A\,\le\,9$ nuclei in units of $\mu_N$ computed for the NV2+3-Ia$^{\star}$ model  and consistent electromagnetic current. The column denoted by TOT is the sum of all contributions to the magnetic moment, while TOT-LO indicates the sub-leading current contribution to the result. The experimental data are from the evaluations of~\cite{Purcell:2010hka,Tilley:2002vg,Tilley:2004zz,Borremans:2005ar}.}
\label{tab:gfmc.mm.light}
\end{table*}

\begin{figure*}[tbh]
\begin{center}
    \includegraphics[width=0.9\textwidth]{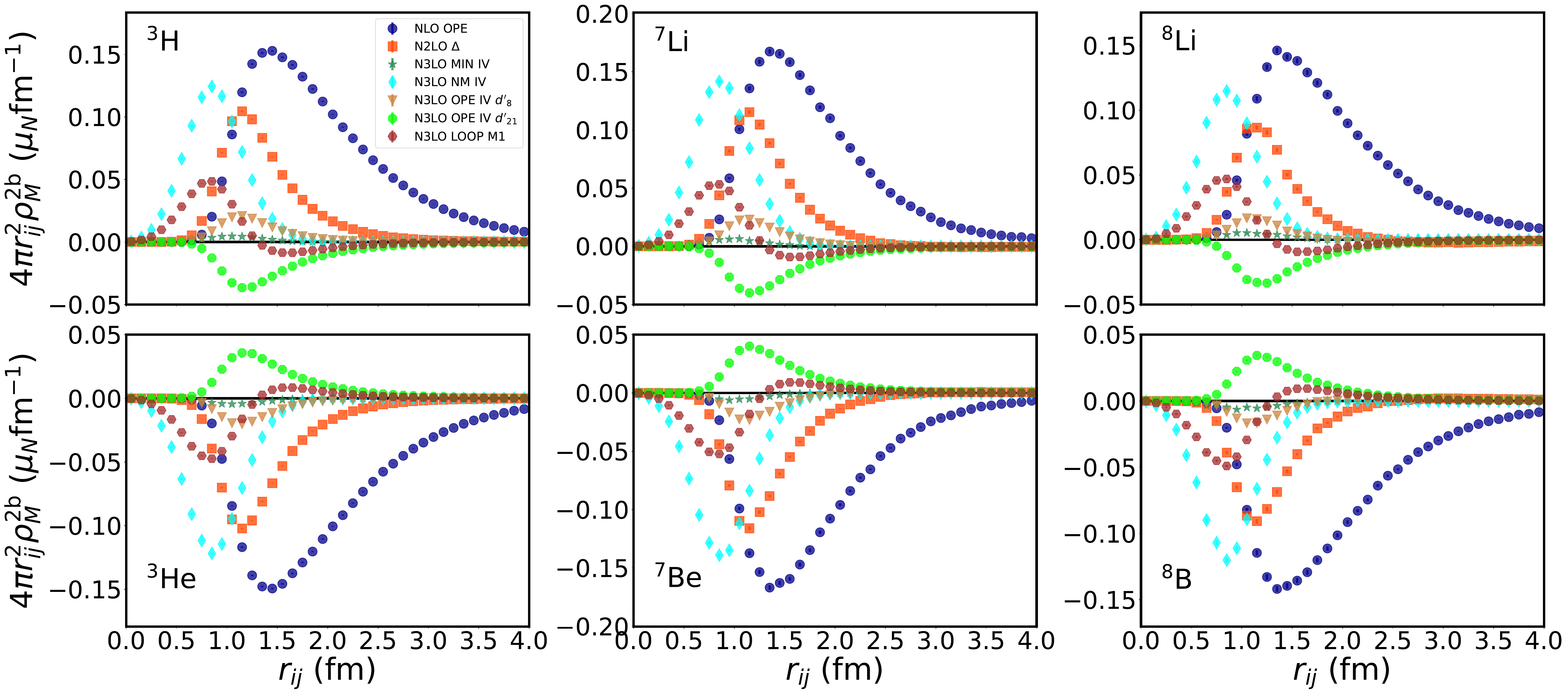}
\end{center}
\caption{VMC calculations of two-body IV magnetic densities for $A\,=\,3,~7,~{\rm and}~8$ systems computed with the NV2+3-IIb$^{\star}$ model as a function of interparticle spacing $r_{ij}$. 
Contributions from the `NLO', `N2LO($\Delta$)', `N3LO OPE IV $d_8^\prime$' and `N3LO OPE IV $d_{21}^\prime$' currents of one-pion range are represented by blue circles, red squares, light brown triangles, and green octagons,  respectively. Contributions from the short-range `N3LO(MIN) IV' and `N3LO(NM) IV' currents are represented by olive green stars and cyan diamonds, respectively; while the TPE IV contribution is indicated by brown hexagons. See text for explanations. 
Statistical uncertainties are included but smaller than the markers representing the central value at a given $r_{ij}$.}
\label{fig:dens.a378}
\end{figure*}

 \begin{figure*}[tbh]
\begin{center}
    \includegraphics[width=0.7\textwidth]{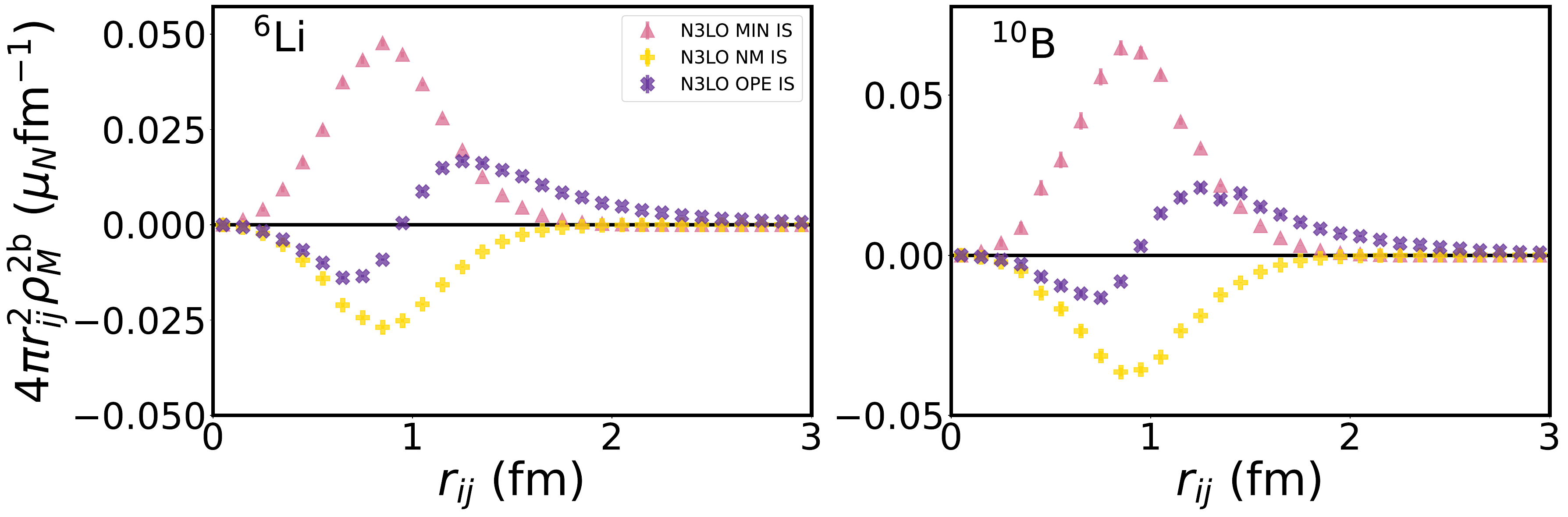}
\end{center}
\caption{VMC calculations of two-body IS magnetic densities for $^6$Li and $^{10}$B computed with the NV2+3-IIb$^{\star}$ model as a function of interparticle spacing $r_{ij}$. 
Contributions from the `N3LO(MIN) IS', `N3LO(NM) IS', and `N3LO(OPE) IS' currents are represented by pink triangles, yellow crosses, and purple times symbols, respectively. See text for explanations.  Statistical uncertainties are included but smaller than the markers representing the central value at a given $r_{ij}$.
}
\label{fig:dens.a610}
\end{figure*}

Given that the $A=7$ nuclei are composed of two symmetric subgroups in the Young Scheme~\cite{Wiringa:2006ih}, having a dominant $\{43\}$ spatial symmetry pattern, there is an $\alpha + {}^3{\rm H}(^3{\rm He})$ cluster structure in $^7{\rm Li}~(^7{\rm Be})$. The nature of that clustering leads to the order-by-order breakdown of their magnetic moments being approximately the same as for $^3\rm H~(^3{\rm He})$. This is because the contribution to the magnetic moment within the $\alpha$ structure ($J_\alpha=0$) is small and any two-body contribution between the clusters is weak due the relatively weak OPE $N\!N$ strength in relative $P$ waves~\cite{Wiringa:2006ih} suppressing meson-exchange contributions, and Pauli blocking suppressing inter-cluster contact contributions. This is made evident by the similarity of the two-body magnetic densities in Fig.~\ref{fig:dens.a378}, defined as
\beq
\mu^{\rm 2b} = \int dr_{ij} \, 4\pi\, r_{ij}^2\, \rho_M^{\rm 2b}(r_{ij})\, .
\eeq
Here, $\mu^{\rm 2b}$ denotes a generic two-body current contribution to the magnetic moment, obtained upon integration of the two-body magnetic density over the interparticle distance $r_{ij}$. In Fig.~\ref{fig:dens.a378}, we plot the IV magnetic densities for $A=3,~7,~{\rm and}~8$ nuclei for model IIb$^{\star}$. In the panels of this figure, the NLO and N2LO($\Delta$) terms of one-pion range, given respectively in Eqs.~(\ref{eq:nlor2}) and~(\ref{eq:jdn2lo}), are represented by blue circles and red squares. The minimal and non-minimal contact currents, N3LO(MIN) IV and N3LO(NM) IV, given in Eqs.~(\ref{eq:jmin.iv}) and~(\ref{eq:jnm.iv}), are indicated with olive green stars and cyan diamonds, respectively. The `N3LO OPE IV $d_8^\prime$' and `N3LO OPE IV $d_{21}^\prime$' one-pion range contributions of Eqs.~(\ref{eq:n3lo.iv.d8}) and~(\ref{eq:n3lo.iv.d21}), are represented by light brown triangles and green octagons, respectively.  Finally, the  LOOP IV contribution at N3LO is represented by brown hexagons, and its expression is given by the IV terms in Eq.~(\ref{eq:jloop}).

Comparing across models, we note that the NLO and N2LO contributions for $A=3$ and $A=7$ exhibit a dependence on the choice of the NV2+3 model. These contributions are found to be more sensitive to changes in the cutoff parameters (class ``a" vs.~class ``b") than to changes in the energy range used to fit the $N\!N$ potential (class ``I" vs.~class ``II"). At N3LO, the N3LO(NM) and N3LO(OPE) contributions show a significant model dependence, which we attribute to the fact that the LECs appearing in these currents are fit to data. In particular, similarly to the findings in Refs.~\cite{Gnech:2022vwr,Martin:2023dhl}, we observe that the N3LO contribution is enhanced relative to naive power counting expectations, largely driven by the N3LO(NM) term which is similar in size to the NLO term. At this order, the other non-minimal term, namely the N3LO(OPE) term, typically provides a contribution consistent in magnitude with N2LO in the power counting, except for model Ia$^{\star}$. For this specific model, a nearly exact cancellation between the two terms entering the N3LO(OPE) IV current--see panel (f) in Fig.~\ref{fig:currn3lo} and Eq.~(\ref{eq:n3lo.iv})--leads to this accidental suppression of the matrix element. This competition is made clear in the plot of  $\rho_M^{2b}$ in Fig.~\ref{fig:dens.a378}. 

Similarly, $^6$Li, with its dominant $\{42\}$ spatial symmetry, displays a strong $\alpha+d$ cluster structure. At leading order, the contribution to the magnetic moment in this system is approximately the value of the deuteron's magnetic moment ($\mu_d$ = 0.857438 $\mu_N$~\cite{Tiesinga:2021myr}), exhibiting minimal model dependence. Further, as $^6$Li is a purely IS ($T=0$) nucleus, the matrix elements for IV components of the current operator vanish. Consequently, some contributions from the current are suppressed beyond what one would expect from power counting considerations. For instance, the NLO contribution to the current is purely IV and thus does not contribute to the magnetic moment of $^6$Li. The N3LO(MIN) and N3LO(NM) contributions in this case are still enhanced, making few percent contributions to the overall magnetic moment. There is a ${\sim}15\%$ model dependence on the N3LO(MIN) contribution that is driven by changing the cutoff, whereas the N3LO(NM) contribution is much more stable across models. 

The N3LO(OPE), contrary to its behavior for $A=3$ and $A=7$, contributes with a size that is consistent with its power counting. The N3LO(OPE) term also displays a large model dependence in $^6$Li, though this term makes a rather small contribution to the overall matrix element. This model dependence is readily understood by looking at Eq.~(\ref{eq:n3lo.is}). As observed in Ref.~\cite{Schiavilla:2018udt}, the N3LO(OPE) IS contribution exhibits a node stemming from the opposite signs of the correlation functions, $I_1(x)$ and $I_2(x)$, appearing in the current. This, combined with the cancellation of the N3LO(MIN) IS  and N3LO(NM) IS  contact terms, leads to an overall small contribution from  two-body IS currents to the total magnetic moment. Similar observations persist when extending our analysis to the two-body IS  magnetic densities across the remaining nuclei examined in this study, which are plotted in Fig.~\ref{fig:dens.a610}. The three IS terms plotted are the `N3LO(OPE) IS,' as well as the `N3LO(MIN) IS'  and `N3LO(NM) IS' contact currents of Eqs.~(\ref{eq:jmin.is}) and~(\ref{eq:jnm.is}). These are represented in the figure with the yellow crosses, pink triangles, and purple ``x" symbols, respectively. 

Finally, the summed magnetic moment for $^6$Li displays very little model dependence. Now, two-body currents only account for ${\sim}2\%$ of the total magnetic moment. 

\subsubsection{VMC magnetic moments of $A=8-10$ nuclei}

Due to the strong cutoff dependence found in the one-pion range terms at NLO and N2LO, the lack of energy range dependence at those orders, and the general model dependence observed in the N3LO terms, we focus on two representative model classes, Ia$^{\star}$ and IIb$^{\star}$, for the heavier nuclei with $A=8-10$ to streamline the computations.

The VMC results for the $A=8-10$ systems are summarized in Table~\ref{tab:vmc.mm.light810}. Referring to this table, we begin our discussion with the case of the purely IS $^{10}$B ground state. Similar to $^6$Li, the IV terms are suppressed relative to their expected power counting. The LO contribution is once again rather stable, albeit differing slightly from that of the deuteron at the same order primarily due to additional contributions from the orbital magnetic moments of the protons. The N3LO(MIN) and N3LO(OPE) contributions display model dependence and have a magnitude that would be consistent with one order lower in the power counting. However, this model dependence cancels out in the summed result, leading to an overall consistency between the two  investigated model classes. As for $^6$Li, two-body currents only play a small role and make up ${\sim}1\%$ of the total magnetic moment.

The addition of an unpaired $p$-shell neutron (proton) to $^7{\rm Li}~(^7{\rm Be})$ forms the predominantly $\{431\}$ spatial symmetry ground state of $^8{\rm Li}~(^8{\rm B})$. At LO, the uncoupled valence nucleon quenches the magnitude of the magnetic moment. Interestingly, the NLO and N2LO contributions are not greatly altered by this additional nucleon. In these nuclei, there is little additional binding from the OPE $N\!N$ interaction with the added nucleon~\cite{Wiringa:2006ih}. Consequently, the dominant one-pion range current contribution at NLO stays roughly the same in $A=8$ as in $A=7$ systems. The similarity is also true for the two-body density, as seen in Fig.~\ref{fig:dens.a378}. This carries over also into the long-range N2LO and N3LO contributions, while the short-range  contributions at N3LO remain enhanced relative to their power counting, and are found to be model dependent. For $^8$B, two-body currents quench the magnetic moment by $\sim30\%$. In $^8$Li, instead, they enhance it by $\sim 20\%$. 

\begin{figure}[tbh]
\begin{center}
    \includegraphics[width=0.46\textwidth]{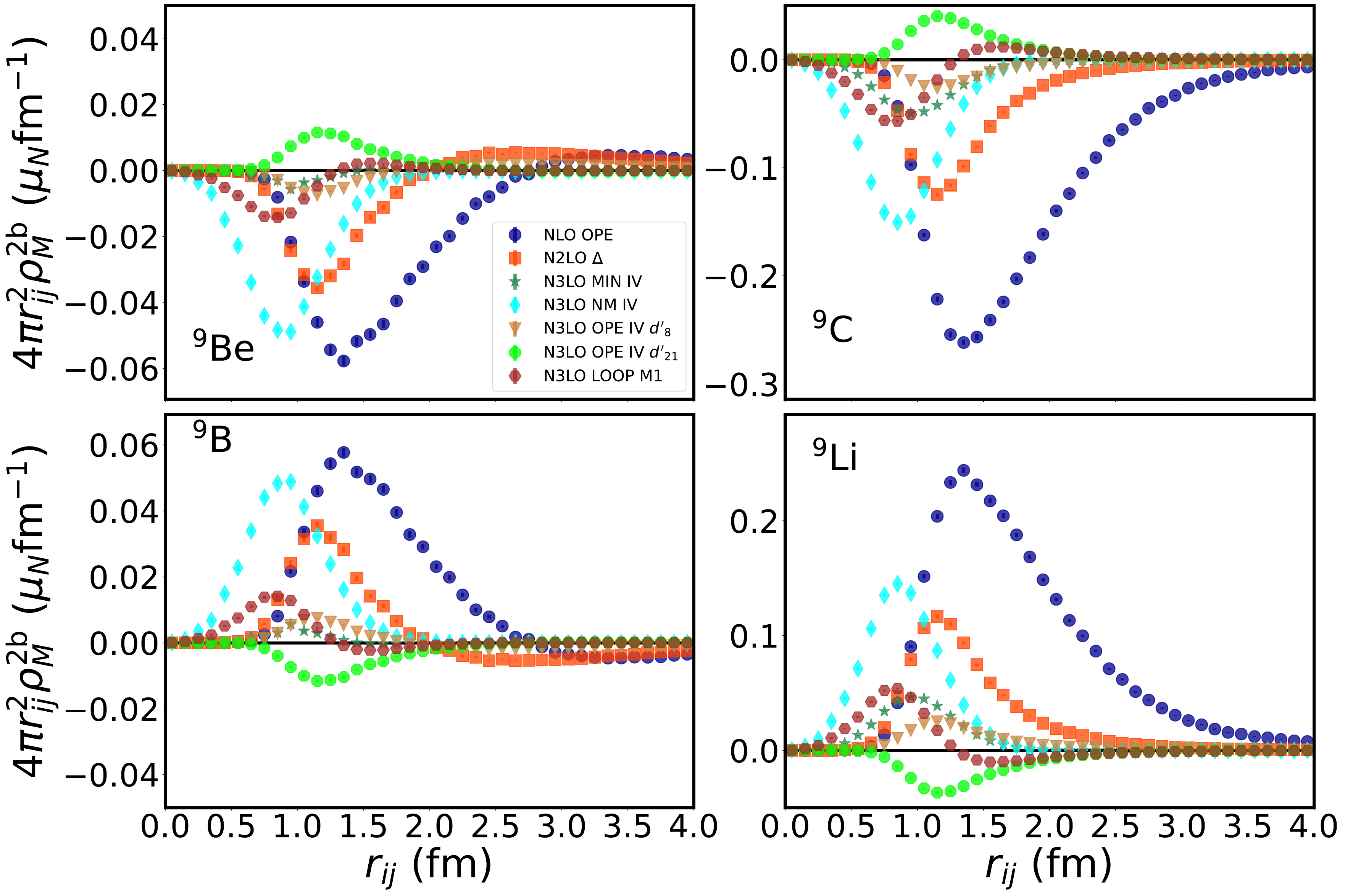}
\end{center}
\caption{Same as Fig.~\ref{fig:dens.a378} but for $A=9$ systems.}
\label{fig:dens.a9}
\end{figure}

For $A=9$ systems, the differences between the $T=3/2$ ground state of $^9$Li and the $T=1/2$ ground state of $^9$Be lead to interesting physics in the two-body currents' effect. As discussed in the previous QMC evaluation within the hybrid approach~\cite{Pastore:2012rp}, the $\{432\}$ spatial symmetry dominates the $^9$Li ground state, giving it a $\alpha + t + 2n$ cluster structure. On the other hand, $^9$Be is dominated by $\{441\}$ spatial symmetry and clusters as $2\alpha + n$. A single valence neutron outside of the $2\alpha$ cluster experiences almost no binding from the OPE $N\!N$ interaction~\cite{Wiringa:2006ih}. Consequently, in this system, the two-body current contribution at NLO is suppressed relative to its power counting. This is reflected in the two-body densities of Fig.~\ref{fig:dens.a9}. 

Before considering mixing from the tensor force, note that the ratio of $p$-shell IS pairs to IV pairs in $^9$Be is enhanced by a factor of 2 relative to $^9$Li. Since the IS pairs will not couple to the IV two-body current at NLO, this brings down its contribution to the magnetic moment in $^9$Be relative to $^9$Li. This suppression, due to purely nuclear structure effects, accounts for the small (${\sim} 6\%$) two-body current contribution to the magnetic moment of $^9$Be. In $^9$Li, instead, there is binding from the OPE $N\!N$ interaction felt both within the triton-cluster and between the triton-cluster and the valence neutrons. Here, two-body currents play a much larger role and make a contribution of ${\sim} 15\%$ to the total calculated magnetic moment. 

The cluster structure also explains the suppression of the contact contributions in $^9$Be relative to $^9$Li. Pauli blocking suppresses contributions between the valence nucleon in $^9$Be and the $\alpha$ clusters. Instead, the two valence nucleons in the $\{432\}$ cluster structure of $^9$Li form a spatially symmetric and IV pair. Thus, the long-range currents contributions are enhanced by this pair over the values for nuclei with $\{43\}$ and $\{431\}$ cluster structures. The spatial symmetry also allows the additional pair of nucleons a non-zero overlap at shorter distances in their relative wave function. This explains the increased N3LO(NM) and N3LO(MIN) contribution in $^9$Li relative to $^8$Li.

The wave functions for the mirror nuclei of $^9$Li and $^9$Be, specifically $^9$C and $^9$B, respectively, were generated by swapping the numbers of protons and neutrons without re-optimizing the variational parameters. Consequently, and for similar reasons, the two-body current contribution in $^9$B is suppressed, while in $^9$C, it significantly impacts the magnetic moment, accounting for $\sim33\%$ of the total calculated value. As observed in the $A\le8$ nuclei, the N3LO contributions in $A=9$ also show dependence on the specific NV2+3 model employed. For $T=3/2$ nuclei, the N3LO(NM) contribution whose magnitude would be more consistent with roughly NLO in the power counting. Comparatively, in $T=1/2$ systems, the N3LO(NM) contribution is at a level that would be more consistent with an N2LO term in the power counting scheme. 

Analyzing the trends in the systematic order-by-order calculations of the magnetic moment  for several nuclei, we see that the N3LO(NM) term tends to be enhanced by roughly two orders for nuclei where the IV currents do not vanish. In the IS case, the enhancement is only by approximately one order. Another standout case is the $A=9,\,T=1/2$ system, where both the NLO(OPE) and N3LO(NM) terms are significantly suppressed due to $2\alpha + N$ cluster structures. 

It is worth noting that the electromagnetic currents of one-pion range entering at NLO and N2LO also vary with the cutoff. An evaluation using two models is certainly not enough to claim a systematic cutoff dependence; however, given the hints of cutoff sensitivity and the enhancement of the non-minimal contact terms that are fit to reproduce $A \leq 3$ experimental data, it is worth rigorously investigating the renormalization of the NLO and N2LO contributions to the magnetic moment. While speculative at this point, it is possible that such an evaluation could indicate the need to promote a contact operator to higher orders to explain the lack of convergence in the power counting observed in this study and in the studies of Refs.~\cite{Gnech:2022vwr,Martin:2023dhl}. Further, the authors of Ref.~\cite{Valderrama:2015} previously argued, on the basis of the wave function anomalous dimension, that the contact operator contributing to the magnetic moment should have a power counting that places it roughly between N2LO and NLO, consistent with the  enhancement observed here and in Refs.~\cite{Gnech:2022vwr,Martin:2023dhl}; however, such an effect could also be due to the implementation of cutoffs that are too soft, causing an uncontrolled sensitivity to the short-range physics~\cite{Hammer:2019poc}. Either case has important implications for the convergence of the power counting and for future studies of electromagnetic observables. As such, studies of the renormalization of simple two-body observables are presently being pursued to further understand these observations and to determine the nature of this enhancement.

\subsubsection{GFMC magnetic moments of $A=3-9$ nuclei} We now turn our attention to the GFMC results summarized in Table~\ref{tab:gfmc.mm.light}. Due to the relatively small model dependence  found in the summed contributions, we perform GFMC propagation only for model Ia$^{\star}$. Typically, we extracted the GFMC result for the magnetic moment by averaging  ${\sim}30$ mixed estimates after convergence is reached. However, for the $^8$B ($^8$Li) ground state we observe a behaviour consistent with the findings of Ref.~\cite{King:2020wmp}. During the GFMC propagation, the point proton (neutron) radius increases monotonically even after convergence is reached for the energy calculation, which typically occurs at $\tau\approx 0.1 {\rm ~MeV}^{-1}$. This may be interpreted as the nucleus dissolving into a $^7$Be ($^7$Li) and a proton (neutron). For these systems,  we  treat $\tau = 0.1 {\rm ~MeV}^{-1}$ as the imaginary time at which spurious contamination has been removed from the VMC wave function, and average for a small number of time steps around this point; namely, in the windows $\tau \in [0.06,0.14]~{\rm MeV}^{-1}$. We estimated the systematic uncertainty by doubling the size of the window to find the change in the average.

For the $A=9, \,T=3/2$ cases, we also have a large uncertainty on the GFMC calculations. In this case, however, the uncertainty is statistical rather than systematic. While the total and beyond leading order results are stable after convergence, there is a large error on these matrix elements. Two matrix elements--namely the NLO and N3LO(MIN)-- have large uncertainties while the sum of the subleading currents is stable. This results in a $5\%$ error for $^9$Li, but for $^9$C, there is a more striking $14\%$ error. In the later case, the LO contribution only has a $3\%$ error but the overall uncertainty is dominated by the two-body term; however, as we treat the uncertainties on the total and total subleading matrix elements as a quadrature sum of the component uncertainties, the summed matrix elements also have a large uncertainty. Of course, this assumes that the errors on the components are uncorrelated, which may not necessarily be the case. Thus, this may be a rather conservative estimate of the total uncertainty. These magnetic moments could make interesting cases to investigate with more rigorous uncertainty quantification and advanced statistical techniques in the future. While immense effort is being carried out in this direction by the nuclear theory community~\cite{Phillips:2020dmw,Furnstahl:2015rha,Melendez:2019izc,Wesolowski:2021cni}, this is beyond the scope of the present work, and as such, we assume the conservative estimate here. 

The GFMC magnetic moments are displayed for $A\leq 9$ nuclei in Fig.~\ref{fig:exp.comp} and are presented in Table~\ref{tab:gfmc.mm.light}. On average, the GFMC propagation changes the VMC magnetic moments by $\lesssim 1.5\%$ for light systems. In heavier nuclei, the change can be more drastic and ranges from $5\%$ to $15\%$. Overall, our calculations agree with experiment on average at the $4\%$ level, ranging from $\lesssim 1\%$ to $7\%$. In this range, only four cases have central values with worse than ${\sim}2\%$ agreement with the experimental data; namely, $^7$Be, $^8$Li, $^8$B, and $^9$Li. The discrepancy for $A=8$ is most likely due to the previously mentioned difficulty modelling these nuclei with model Ia$^{\star}$, as the chiral potential tends to be unstable to breakup into an $A=7$ nucleus plus a free nucleon. For the remaining cases, the change from VMC to GFMC is driven primarily by the one-body term, which is most sensitive to the nucleon spin distributions. Since the VMC already provided extremely good agreement with the data in these cases, it may be unsurprising that GFMC propagation could produce worse agreement, since it alters the spin-spin correlations in the system. We stress that while the GFMC provides the most accurate calculation within a model, this does not imply it will provide better agreement with data than the VMC. Thus, further attention is merited to understand the inability of the Norfolk model to obtain few percent agreement with the data in these specific cases and to improve nuclear models based on $\chi$EFT moving forward. 

We note that two-body currents still play a large role in the total GFMC magnetic moments for all but the IS nuclei. As expected, the overall two-body correction still provides a small change to the LO result. In fact, recent many-body calculations of magnetic moments in heavy mass nuclei retaining only the first subleading correction at NLO have highlighted that two-body contributions remain important in heavy systems~\cite{Miyagi:2023zvv}. Finally, in some cases, the two-body current contribution can change sign. These cases all correspond to currents whose two-body densities show the presence of a node due to an interplay of operators with opposite signs.

\subsubsection{Scaled two-body magnetic densities}

\begin{figure}[tbh]
\begin{center}
    \includegraphics[width=0.46\textwidth]{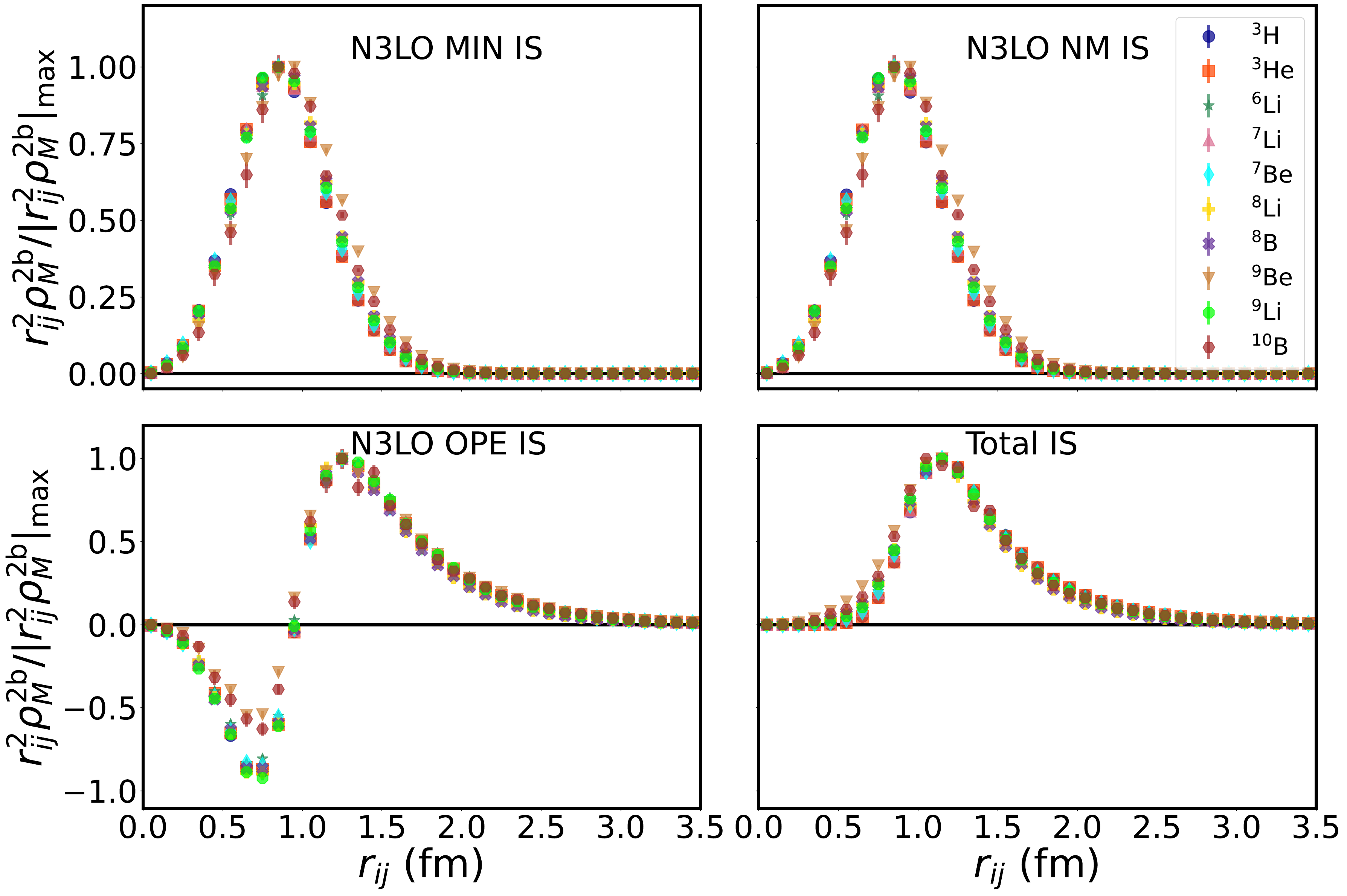}
\end{center}
\caption{Scaled two-body VMC IS magnetic densities for different nuclei computed with model IIb$^{\star}$ as a function of interparticle spacing $r_{ij}$. Different colors represent different nuclei.}
\label{fig:dens.is}
\end{figure}

\begin{figure*}[tbh]
\begin{center}
    \includegraphics[width=0.9\textwidth]{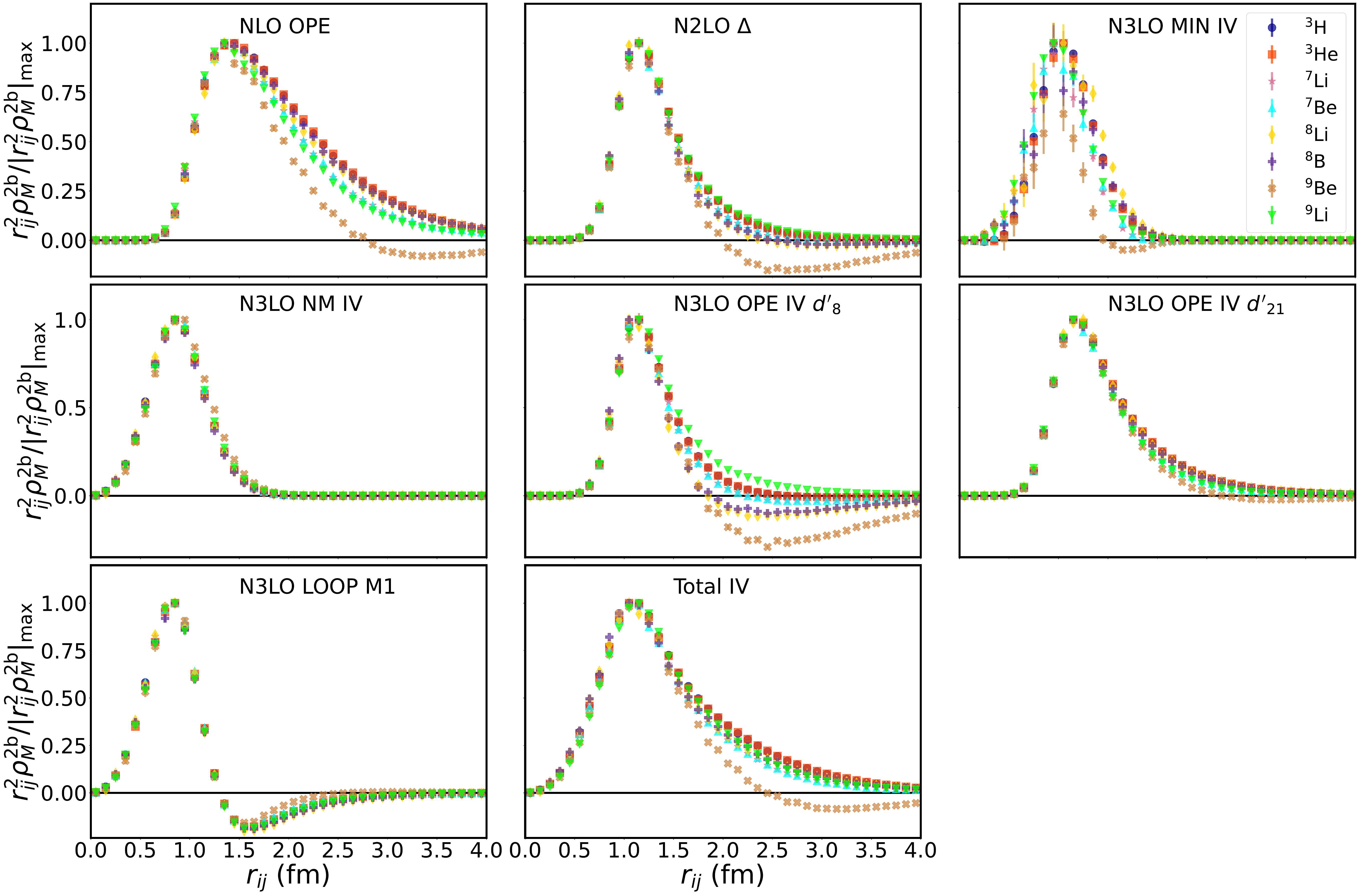}
\end{center}
\caption{Scaled two-body VMC IV magnetic densities for different nuclei computed with model IIb$^{\star}$ as a function of interparticle spacing $r_{ij}$. Different colors represent different nuclei.}
\label{fig:dens.iv}
\end{figure*}

The study of scaled two-body magnetic densities is relevant to identify universal behaviours in both short- and long-range two-nucleon dynamics. Such universal behaviors can be understood in terms of generalized contacts~\cite{Cruz-Torres:2019fum} or from RG evolution~\cite{Tropiano:2021qgf}. In a similar spirit, to examine the shape of the two-body magnetic densities, we scale them by the maximum value of $|r_{ij}^2\rho^{2b}_M|$, and analyze  the scaled densities, defined as $r_{ij}^2\rho^{2b}_M/|r_{ij}^2\rho^{2b}_M|$. In Figs.~\ref{fig:dens.is} and~\ref{fig:dens.iv}, we plot the scaled IS and IV two-body magnetic densities, respectively, for all the nuclei considered in the present work.

Referring to these figures, we note that the scaled IS magnetic densities display a striking universal behavior across $r_{ij}$. On the other hand, the scaled IV densities are universal at short range and display a tail which depends on the particular nucleus under study. In Ref.~\cite{King:2020wmp}, a similar behavior was observed for axial transition densities in a study of Gamow-Teller $\beta$-decay matrix elements. The explanation for the universal behavior of the magnetic densities follows a similar logic. At short distances ($r_{ij} \lesssim 1/m_{\pi}$), there are very few pairs in relative $P$-waves due to the repulsive nature of the nuclear force in odd partial waves and the centrifugal barrier. Thus, IS pairs at short distances tend to be predominantly $ST=10$ and IV pairs $ST=01$. Both the $ST=01$ and $ST=10$ pair distributions in nuclei have similar shapes and differ only by an overall scaling factor~\cite{Forest:1996kp,Cruz-Torres:2019fum}. Therefore, this universal behavior is a result of the short-range correlations in nuclei that are driven by the tensor component of the nuclear force~\cite{Schiavilla:2006xx}. At larger distances, since there is less suppression of $P$-wave pairs, and because the $ST=00$ and $11$ pair distributions do not scale, the two-body magnetic density is sensitive to the structure of the ground state; however, because $ST=00$ pairs are formed rather infrequently in light systems~\cite{Wiringa:2006ih,Piarulli:2022ulk}, the long-range tail of the N3LO(OPE) IS  term does not strongly deviate from universal behavior. Instead, due to tensor correlations in the wave function, it is possible to form a large number of $ST=11$ pairs in light systems, which spoils universality for the total IV term at long distances. The near universal behavior of the total IS contribution, the short range universality of the IV term, and the long-range nucleus dependence of the IV term are well understood on the basis of pair formation inside of the nucleus.

\subsubsection{Model dependencies in two-body magnetic densities}

\begin{figure}[tbh]
\begin{center}
    \includegraphics[width=0.45\textwidth]{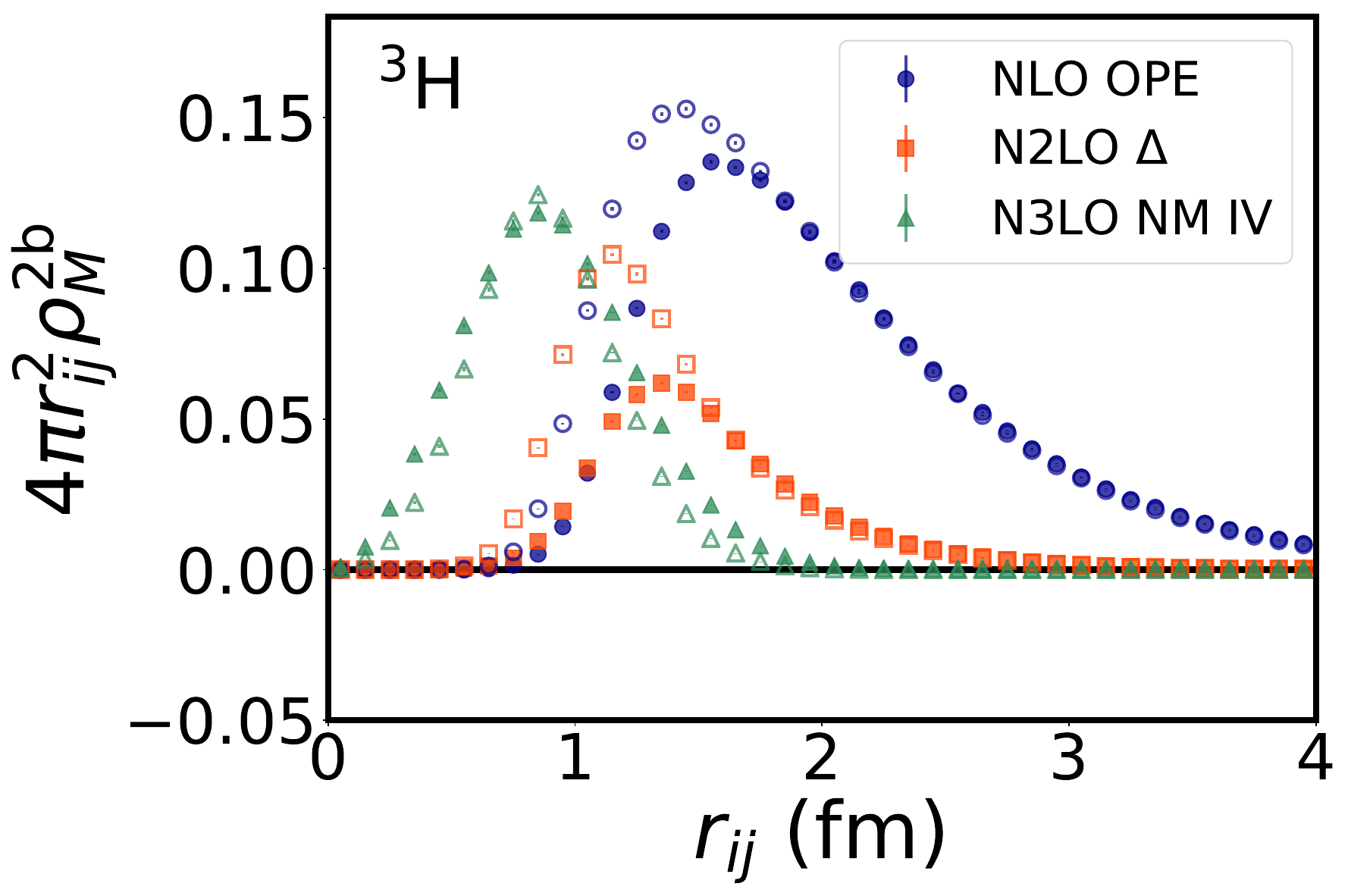}
\end{center}
\caption{Two-body VMC magnetic densities arising from the NLO (blue circles), N2LO-$\Delta$ (orange squares), and the N3LO(NM) IV (green triangles) contributions for $^3$H computed with model Ia$^{\star}$ (filled symbols) and IIb$^{\star}$ (open symbols) as a function of interparticle spacing $r_{ij}$. }
\label{fig:dens.comp.iv}
\end{figure}

\begin{figure}[tbh]
\begin{center}
    \includegraphics[width=0.45\textwidth]{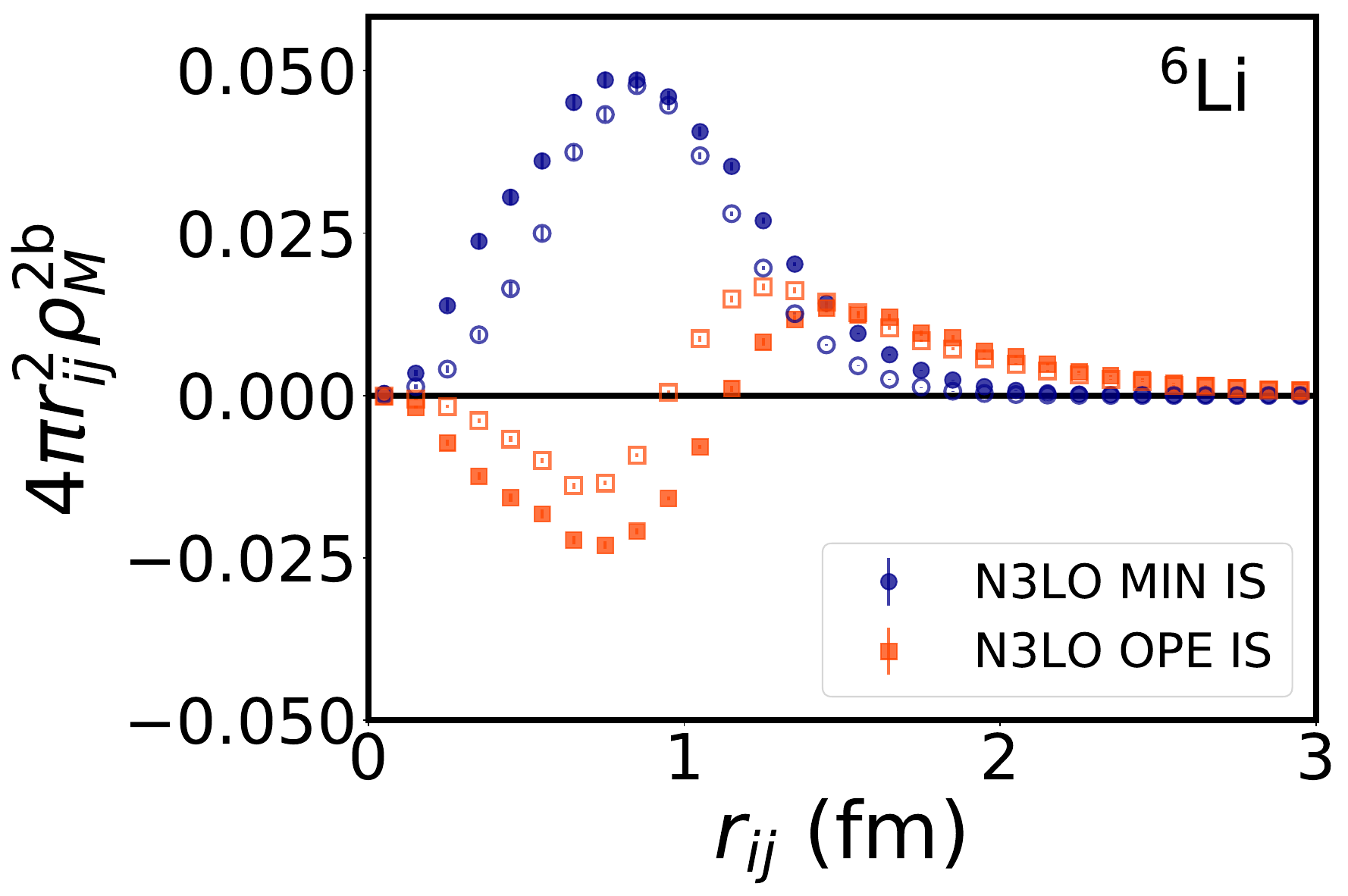}
\end{center}
\caption{Two-body VMC magnetic densities arising from the N3LO(MIN) IS (blue circles), and the N3LO(OPE) IS (orange squares) contributions for $^6$Li computed with model Ia$^{\star}$ (filled symbols) and IIb$^{\star}$ (open symbols) as a function of interparticle spacing $r_{ij}$.}
\label{fig:dens.comp.is}
\end{figure}

Up to this point, we have analyzed the densities only for one model; however, it is important to assess the cutoff dependence, which is evidently important in the integrated two-body contributions. To that end, we compare the dominant IV contributions to the $^3$H magnetic moment in Fig.~\ref{fig:dens.comp.iv}. The effect of the softer cutoff in the Ia$^{\star}$ regulator on the NLO(OPE) and N2LO($\Delta)$ currents is, unsurprisingly, to reduce the two-body magnetic density at short distances; however, the tail of the density is unaffected by the cutoff function. Thus, taking a softer cutoff quenches the overall magnitude of the matrix element. This particularly impacts the N2LO($\Delta)$ current which is of shorter-range than the NLO contribution due to the presence of the intermediate $\Delta$-isobar. The soft cutoff removes strength from the N2LO($\Delta$) magnetic moment around its peak, resulting in a large quenching of the contribution in model Ia$^{\star}$ relative to model IIb$^{\star}$. The N3LO(NM) contribution, however, is enhanced when using the softer cutoff of model Ia$^{\star}$. The contact density is broadened for the softer cutoff while preserving the overall magnitude. This effect is likely due to a combination of having a broader Gaussian representation of the $\delta$-function and a larger density of short-range IV pairs. 

A similar broadening of the contact densities is observed in Fig.~\ref{fig:dens.comp.is} which shows the IS densities plotted for $^6$Li. For the long-range N3LO(OPE) IS contribution, the correlation functions $I_1(x)$ and $I_2(x)$ are particularly sensitive to the choice of regulator, as was previously observed in Ref.~\cite{Schiavilla:2018udt}. This can move the node in the density and change the overall sign of the contribution to the magnetic moment from this current, as shown in the figure.

From the model comparison, it is clear that altering the short-range dynamics with different cutoffs significantly impacts two-body magnetic moments. Clearly, if one takes a cutoff that is too small, it will spoil the convergence behavior of the higher-order contributions due to significant quenching of the NLO and N2LO($\Delta)$ terms. This quenching is compensated for by a larger N3LO(NM) IV fit contribution and thus is not seen if one looks only at the total magnetic moment; however, if one were to perform the calculation up to NLO in the current, evidently the model dependence would induce an error much larger than the naive expectation from power counting. Given this significant model dependence of the two-body contributions and the particular importance of the IV piece in predicting magnetic structure, it is crucial to investigate this cutoff dependence more rigorously in the future. 

\subsection{Magnetic form factors and high-momentum structure}
\label{sec:ff}

\begin{figure*}
   \centering
    \begin{subfigure}[t]{0.28\textwidth}
        \centering
        \includegraphics[width=\linewidth]{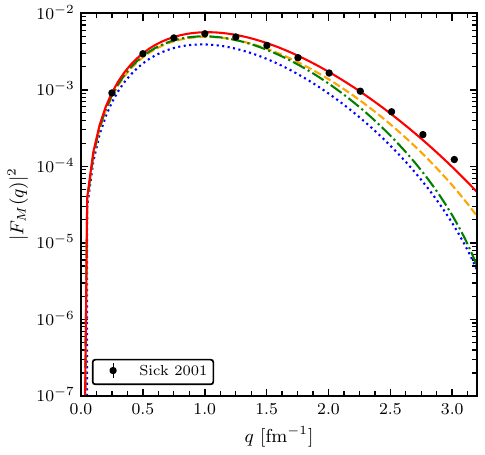} 
        \caption{$\tri$}
    \end{subfigure}
    \begin{subfigure}[t]{0.28\textwidth}
        \centering
        \includegraphics[width=\linewidth]{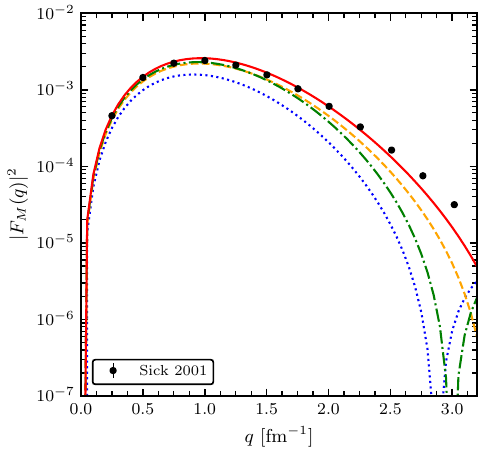} 
        \caption{$\het$}
    \end{subfigure}
        \begin{subfigure}[t]{0.28\textwidth}
        \centering
        \includegraphics[width=\linewidth]{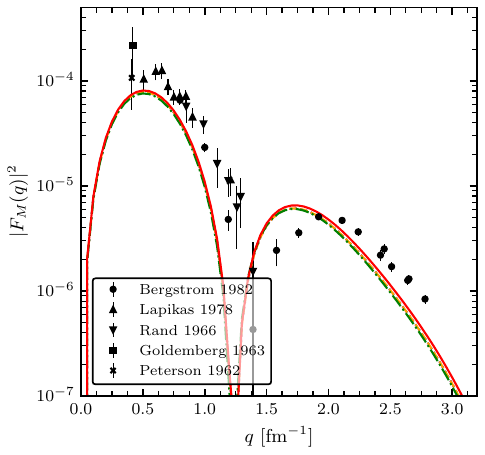} 
        \caption{$\lisix$}
    \end{subfigure}\\
       \begin{subfigure}[t]{0.28\textwidth}
        \centering
        \includegraphics[width=\linewidth]{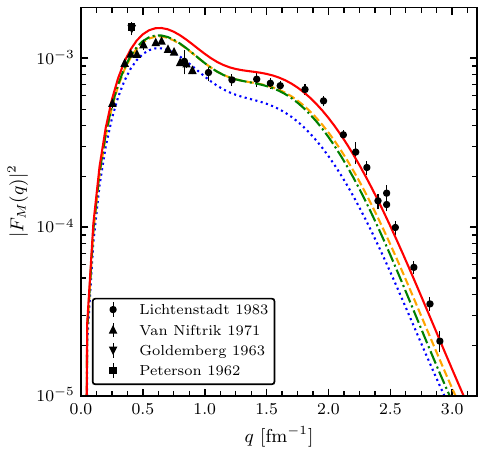} 
        \caption{$\lisev$}
    \end{subfigure}
    \begin{subfigure}[t]{0.28\textwidth}
        \centering
        \includegraphics[width=\linewidth]{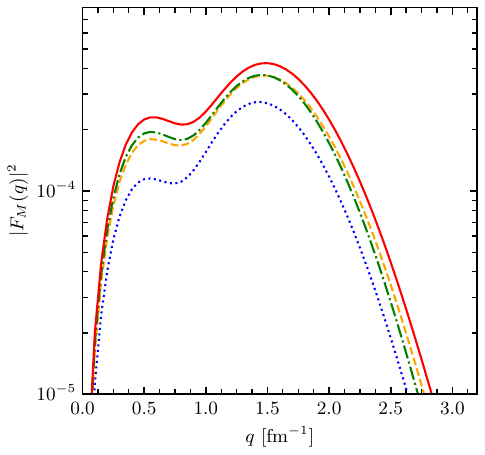} 
        \caption{$\besev$}
    \end{subfigure}
        \begin{subfigure}[t]{0.28\textwidth}
        \centering
        \includegraphics[width=\linewidth]{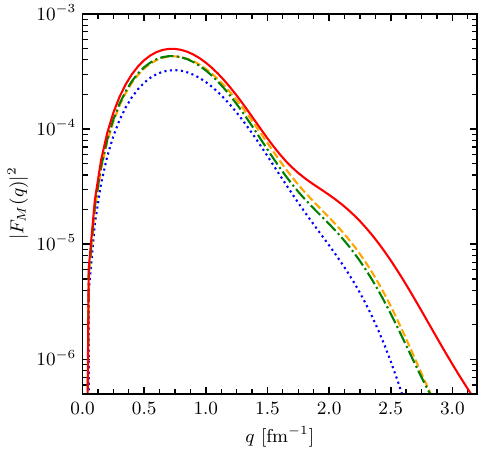} 
        \caption{$\lieight$}
    \end{subfigure}\\
      \begin{subfigure}[t]{0.28\textwidth}
        \centering
        \includegraphics[width=\linewidth]{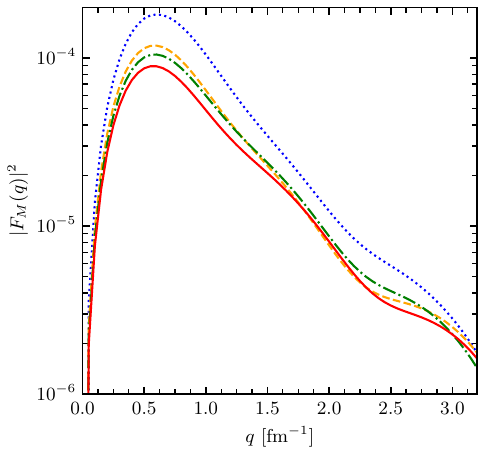} 
        \caption{$\beight$}
    \end{subfigure}
    \begin{subfigure}[t]{0.28\textwidth}
        \centering
        \includegraphics[width=\linewidth]{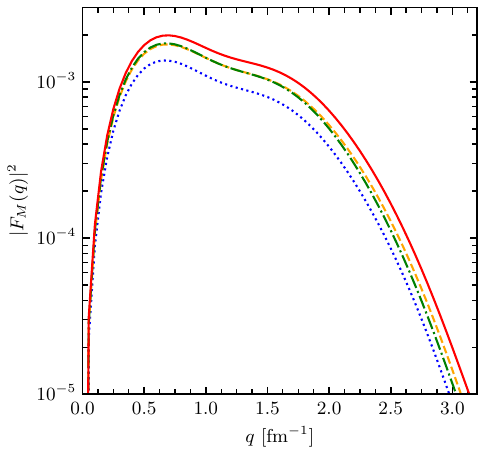} 
        \caption{$\linin$}
    \end{subfigure}
        \begin{subfigure}[t]{0.28\textwidth}
        \centering
        \includegraphics[width=\linewidth]{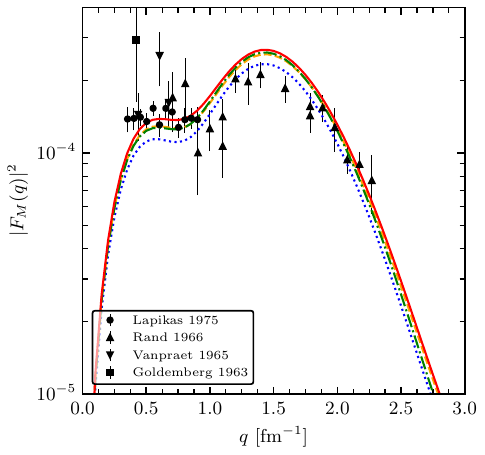} 
        \caption{$\benin$}
    \end{subfigure}\\
          \begin{subfigure}[t]{0.28\textwidth}
        \centering
        \includegraphics[width=\linewidth]{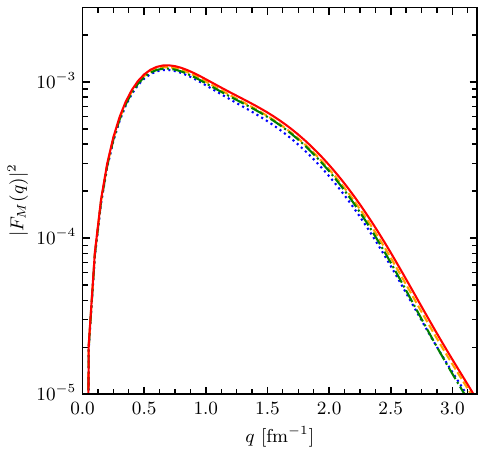} 
        \caption{$\bnin$}
    \end{subfigure}
    \begin{subfigure}[t]{0.28\textwidth}
        \centering
        \includegraphics[width=\linewidth]{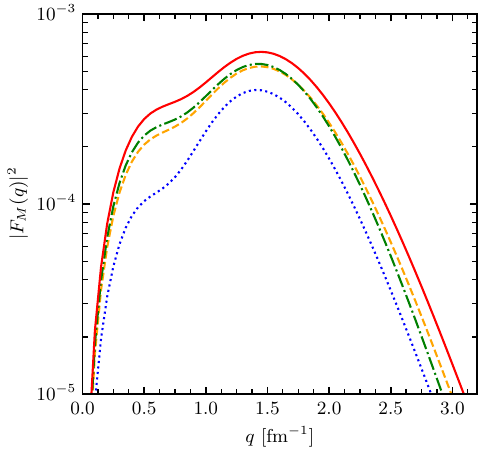} 
        \caption{$\cnin$}
    \end{subfigure}
        \begin{subfigure}[t]{0.28\textwidth}
        \centering
        \includegraphics[width=\linewidth]{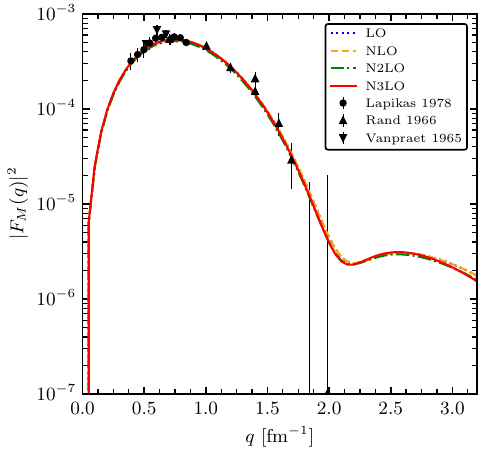} 
        \caption{$\bten$}
    \end{subfigure}
    \caption{Magnetic form factors computed with the NV2+3-IIb$^{\star}$ interaction and consistent electromagnetic current. In the figure we show the contribution of the currents up to LO (dotted blue), NLO (dashed yellow), N2LO (dot dashed green) and N3LO (full red) in the chiral expansion. We compare our calculations with the available world experimental data. See the text for more details.}\label{fig:MFF_IIb_co}
\end{figure*}

\begin{figure*}
   \centering
       \begin{subfigure}[t]{0.32\textwidth}
        \centering
        \includegraphics[width=\linewidth]{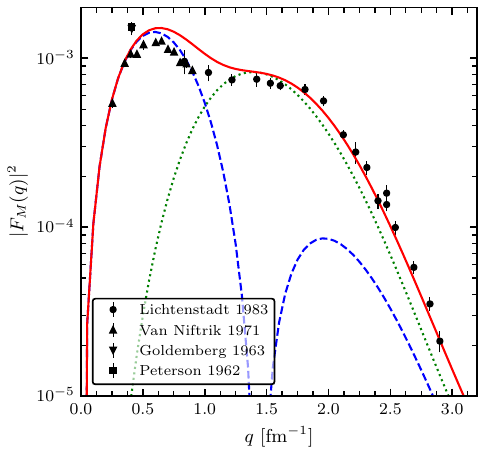} 
        \caption{$\lisev$}
    \end{subfigure}
    \begin{subfigure}[t]{0.32\textwidth}
        \centering
        \includegraphics[width=\linewidth]{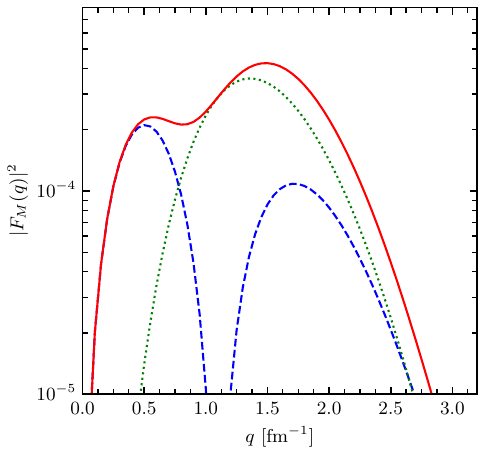} 
        \caption{$\besev$}
    \end{subfigure}
        \begin{subfigure}[t]{0.32\textwidth}
        \centering
        \includegraphics[width=\linewidth]{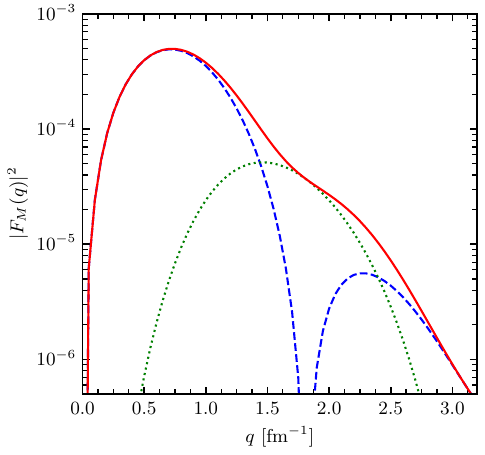} 
        \caption{$\lieight$}
    \end{subfigure}\\
      \begin{subfigure}[t]{0.32\textwidth}
        \centering
        \includegraphics[width=\linewidth]{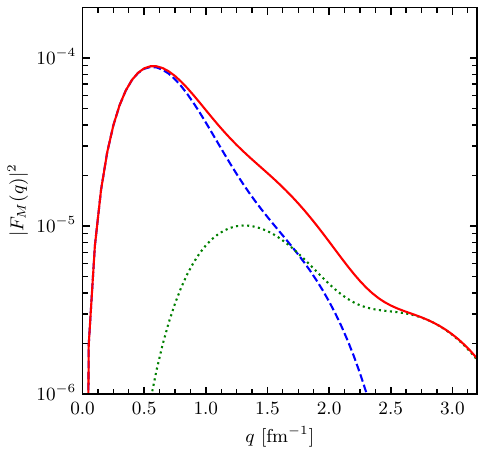} 
        \caption{$\beight$}
    \end{subfigure}
    \begin{subfigure}[t]{0.32\textwidth}
        \centering
        \includegraphics[width=\linewidth]{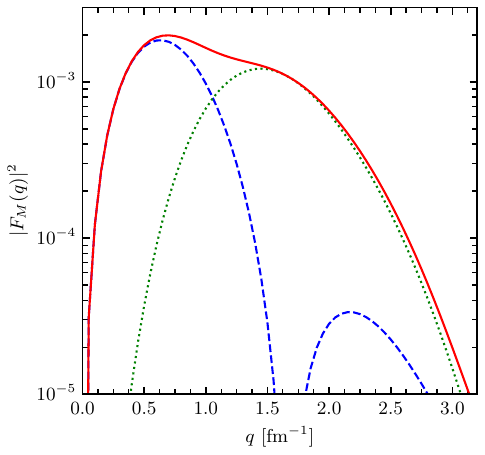} 
        \caption{$\linin$}
    \end{subfigure}
        \begin{subfigure}[t]{0.32\textwidth}
        \centering
        \includegraphics[width=\linewidth]{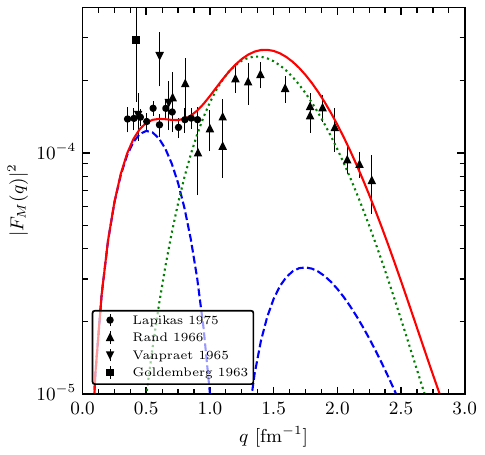} 
        \caption{$\benin$}
    \end{subfigure}\\
          \begin{subfigure}[t]{0.32\textwidth}
        \centering
        \includegraphics[width=\linewidth]{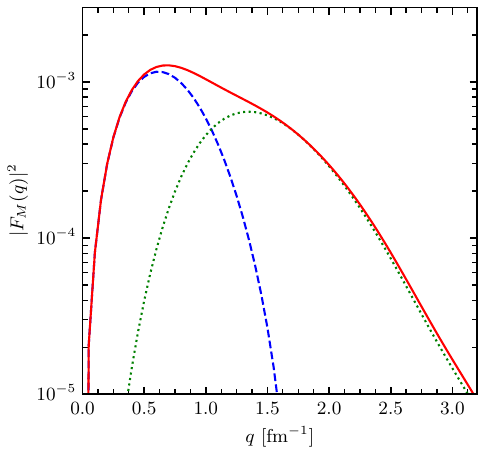} 
        \caption{$\bnin$}
    \end{subfigure}
    \begin{subfigure}[t]{0.32\textwidth}
        \centering
        \includegraphics[width=\linewidth]{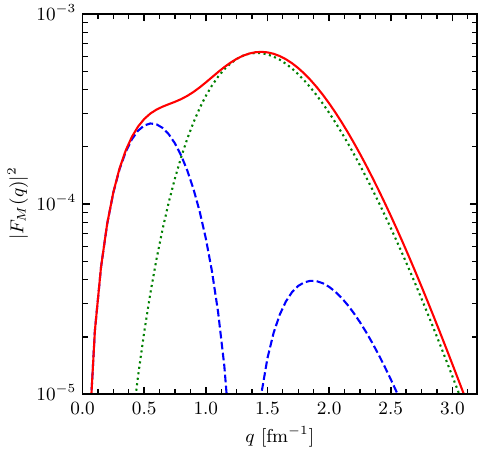} 
        \caption{$\cnin$}
    \end{subfigure}
        \begin{subfigure}[t]{0.32\textwidth}
        \centering
        \includegraphics[width=\linewidth]{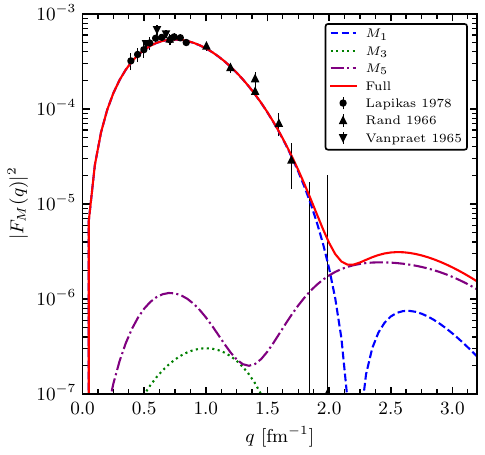} 
        \caption{$\bten$}
    \end{subfigure}
    \caption{Multipole contribution to the form factors computed with the NV2+3-IIb$^*$ interaction and consistent electromagnetic current at N3LO for nuclei with $J>1$. In the figure we show the contribution of the multipoles $M_1$ (dashed blue line), $M_3$ (dotted green), and $M_5$ (dashed dot purple). The red line represents the full result.}\label{fig:MFF_multipole_IIb}
    \end{figure*}

The magnetic form factor has been computed using the same interactions and currents used for the magnetic moments. In Fig.~\ref{fig:MFF_IIb_co}, we present the results for all the nuclei showing the contributions of the currents order-by-order for the NV2+3-IIb$^\star$ interaction compared to data where available. We also include predicted magnetic form factors of radioactive isotopes, because while there are presently no data for comparison, novel techniques have recently made it possible to study electron scattering from radioactive nuclei~\cite{Tsukada:2023}. 

In Fig.~\ref{fig:MFF_multipole_IIb} we present the multipole contributions to the magnetic form factors for the nuclei with $J>1$ computed with the full currents at N3LO and the NV2+3-IIb$^\star$. Similar results are obtained for the other interaction and current models considered.

Finally, we estimated the truncation error, limiting our analysis to the error associated with the chiral expansion of the electromagnetic currents, assuming the power counting adopted in Refs.~\cite{Pastore:2008ui,Pastore:2009is,Pastore:2011ip,Piarulli:2012bn,Schiavilla:2018udt,Gnech:2022vwr}. The calculation of the full chiral truncation error would require an analysis of the chiral expansion in both the interaction and currents, together; however, for the nuclear interactions considered in this work, this is not presently possible.

In order to estimate the truncation error we use the prescription
of Ref.~\cite{Epelbaum:2014efa,Epelbaum:2014sza} applied to
\begin{equation}\label{eq:mff_tilde}
    \tilde F_M(q)=\sqrt{2\pi(2J+1)}\frac{2m_N}{q}\langle J J,J -\!J |1 0\rangle F_M(q)\,.
\end{equation}
 For this analysis, we select a value of $\Lambda_b=700$ MeV and, for the characteristic scale of momentum $Q$, we follow Ref.~\cite{Philips2016} by setting it to the typical momentum transfer $(A-1)/A\,q$, for the nucleus of mass number $A$.  
In Fig.~\ref{fig:MFF_err}, we plot the error bands at LO (orange), NLO (green), N2LO (blue) and N3LO(red). Note that  we do not renormalize the results to a specific value of the magnetic moment.

Below we discuss in detail the results obtained for each nucleus studied.

\subsubsection{Tritium and Helium-3}

The experimental magnetic form factors (shortened to f.f.'s, throughout the reminder of this section) of $\tri$
and $\het$~\cite{Amroun:1994qj, Sick2001} are well reproduced by the VMC calculation as can be seen in Figs.~\ref{fig:MFF_IIb_co}a and~\ref{fig:MFF_IIb_co}b.  The contribution of the N3LO(NM) and N3LO(OPE) fitted terms are able to fill in the diffraction minimum appearing at $q\simeq3$ fm$^{-1}$ in the magnetic f.f.'s of $\het$. The $\tri$ diffraction is not visible in the figure but a similar effect is present. We refer to Ref.~\cite{Gnech:2022vwr} for a more detailed discussion. 
Similar results are obtained for all the other NV2+3 interactions.
Note that the VMC calculations are in perfect agreement with the Hyperspherical Harmonics results of Ref.~\cite{Gnech:2022vwr}.

\subsubsection{Lithium-6}

\begin{figure*}
   \centering
    \begin{subfigure}[t]{0.28\textwidth}
        \centering
        \includegraphics[width=\linewidth]{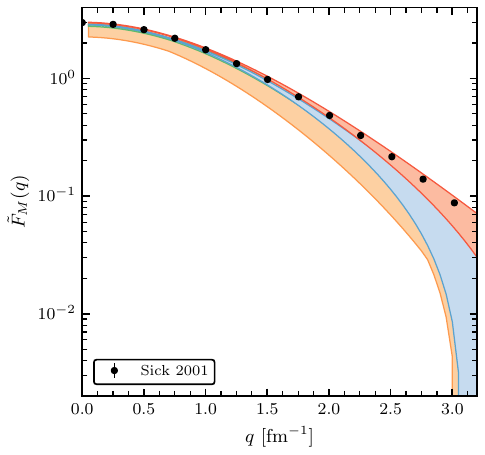} 
        \caption{$\tri$}
    \end{subfigure}
    \begin{subfigure}[t]{0.28\textwidth}
        \centering
        \includegraphics[width=\linewidth]{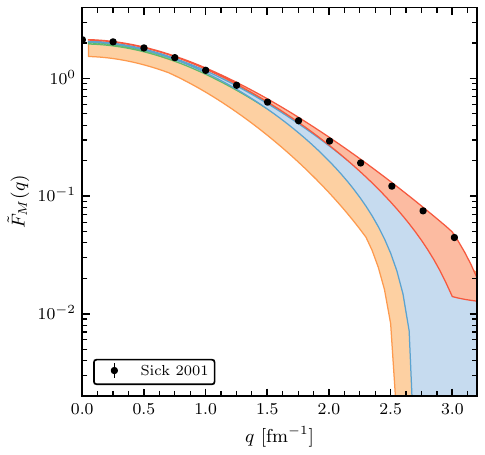} 
        \caption{$\het$}
    \end{subfigure}
        \begin{subfigure}[t]{0.28\textwidth}
        \centering
        \includegraphics[width=\linewidth]{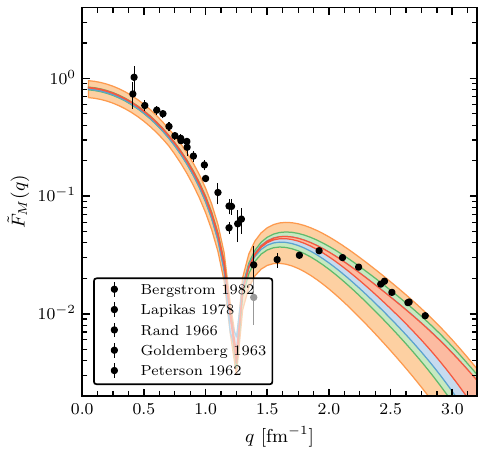} 
        \caption{$\lisix$}
    \end{subfigure}\\
       \begin{subfigure}[t]{0.28\textwidth}
        \centering
        \includegraphics[width=\linewidth]{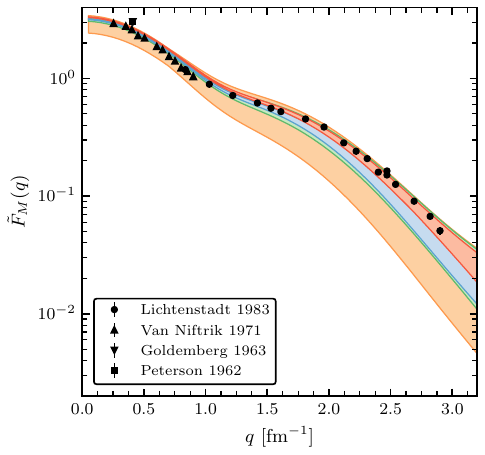} 
        \caption{$\lisev$}
    \end{subfigure}
    \begin{subfigure}[t]{0.28\textwidth}
        \centering
        \includegraphics[width=\linewidth]{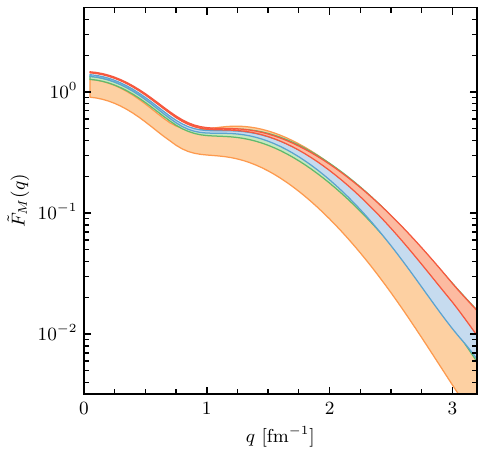} 
        \caption{$\besev$}
    \end{subfigure}
        \begin{subfigure}[t]{0.28\textwidth}
        \centering
        \includegraphics[width=\linewidth]{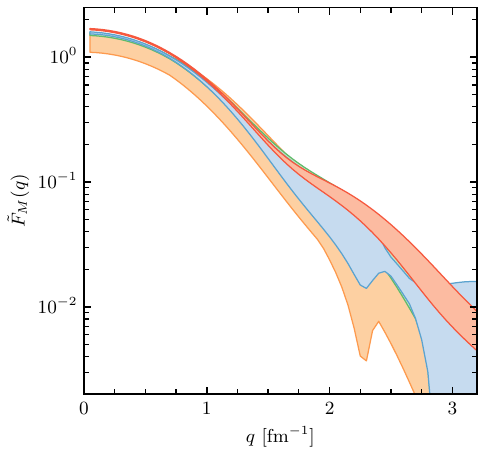} 
        \caption{$\lieight$}
    \end{subfigure}\\
      \begin{subfigure}[t]{0.28\textwidth}
        \centering
        \includegraphics[width=\linewidth]{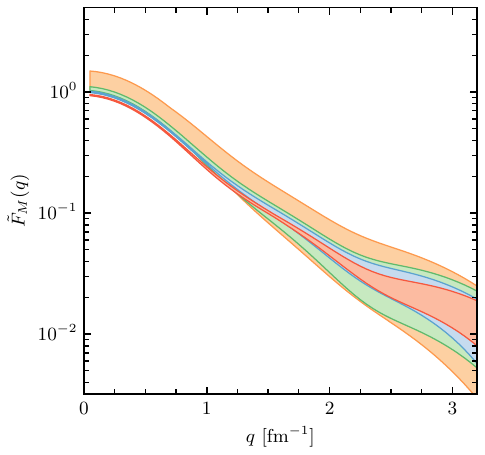} 
        \caption{$\beight$}
    \end{subfigure}
    \begin{subfigure}[t]{0.28\textwidth}
        \centering
        \includegraphics[width=\linewidth]{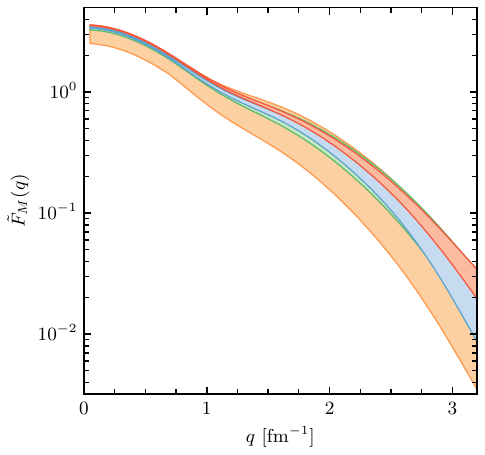} 
        \caption{$\linin$}
    \end{subfigure}
        \begin{subfigure}[t]{0.28\textwidth}
        \centering
        \includegraphics[width=\linewidth]{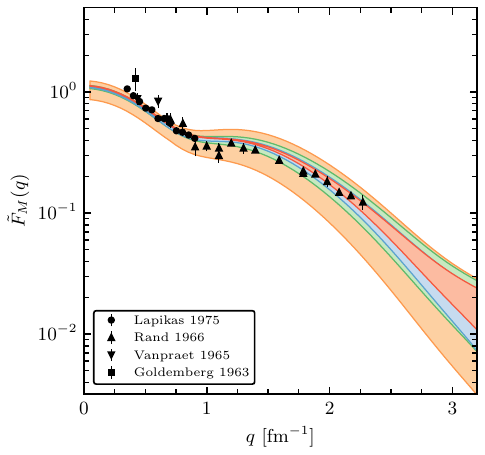} 
        \caption{$\benin$}
    \end{subfigure}\\
          \begin{subfigure}[t]{0.28\textwidth}
        \centering
        \includegraphics[width=\linewidth]{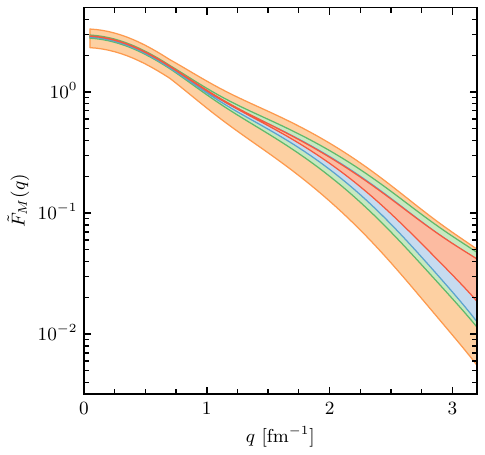} 
        \caption{$\bnin$}
    \end{subfigure}
    \begin{subfigure}[t]{0.28\textwidth}
        \centering
        \includegraphics[width=\linewidth]{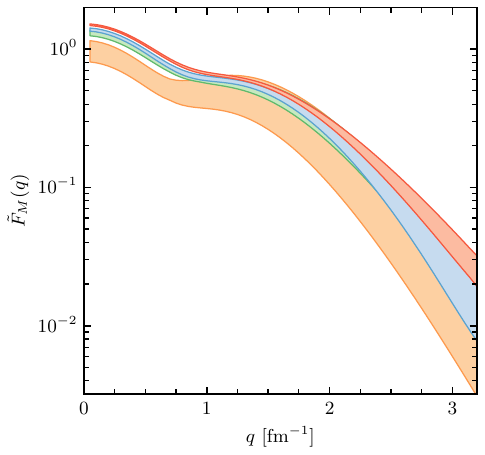} 
        \caption{$\cnin$}
    \end{subfigure}
        \begin{subfigure}[t]{0.28\textwidth}
        \centering
        \includegraphics[width=\linewidth]{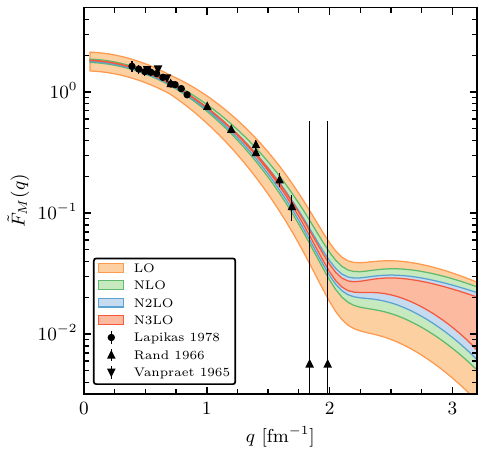} 
        \caption{$\bten$}
    \end{subfigure}
    \caption{Errors generated by the truncation of the chiral current (only) for the magnetic form factors computed with the NV2+3-IIb$^*$ interaction. In the figure the bands represent the error computed at  LO (orange), NLO (green), N2LO (blue) and N3LO (red). Note that for this figure we used the definition of the form factor as in Eq.~\ref{eq:mff_tilde}.}\label{fig:MFF_err} 
\end{figure*}

 The $\lisix$ experimental magnetic form factor presents two peaks at $q\simeq0.5$ and 2 fm$^{-1}$~\cite{Peterson1962,Goldemberg1963, Rand1966,Lapikas1978,Bergstrom1982}. Our calculations are able to reproduce the two peaks even if, for all the interactions considered, the zero tends to come a little too early,  and the second peak contains too much strength. This is evidenced when comparing the theoretical error bands with the experimental data in Fig.~\ref{fig:MFF_err}c. This is in line with the results of Ref.~\cite{Wiringa:1998hr} obtained using  VMC together with phenomenological interactions and currents; however, the phenomenological interactions are able to reproduce the strength of the first peak~\cite{Wiringa:1998hr}, while all the models used in this work underpredict the strength of the first peak (see Fig.~\ref{fig:MFF_IIb_co}c). Since $^6$Li is a $T=0$ nucleus, only two-body IS currents contribute to the magnetic f.f.. These contributions are negligible in the low $q$ region (see Ref.~\cite{Donnelly1984,Wiringa:1998hr}) and become sizable only at large values of $q$; indeed the IS corrections at N3LO are visible only for $q\gtrsim1.5$ fm$^{-1}$, seen in Fig.~\ref{fig:MFF_IIb_co}c. 

\subsubsection{Lithium-7 and Beryllium-7} 

The $\lisev$ and $\besev$ magnetic f.f.'s receive contributions from $M_1$ and $M_3$ multipoles, as can be seen in Figs.~\ref{fig:MFF_multipole_IIb}a and~\ref{fig:MFF_multipole_IIb}b. The $M_3$ contribution becomes dominant in the region $1.0\leq q \leq 2.5$ fm$^{-1}$ where it fills in the diffraction minimum appearing in $M_1$. This is consistent with the findings of Ref.~\cite{Donnelly1984} and references there in, where $\lisev$ is analyzed using phenomenological models. The $M_1$ and $M_3$ contributions are different in the two mirror nuclei. In $\besev$, $M_3$ has more strength than $M_1$, in contrast to what is observed in $\lisev$.

The available data for $\lisev$ are obtained from Refs.~\cite{Peterson1962,Goldemberg1963,Vanniftrik1971, Lichtenstadt1983}. As can be seen in Fig.~\ref{fig:MFF_IIb_co}d, we are able to explain the shape and magnitude of the experimental magnetic f.f.'s.
The cumulative contribution from LO and NLO currents is sufficient to reproduce the peak at $q\simeq0.5$ fm$^{-1}$. 
Conversely, the region where $q\geq 1$ fm$^{-1}$ is significantly underestimated, unless N3LO currents are incorporated. The N3LO contributions are required to describe the tail of the magnetic f.f.; however, they generate too much strength for the $M_1$ peak, despite accurately predicting the magnetic moment. In the region with $q\geq 1$ fm$^{-1}$, the magnetic f.f. is reasonably reproduced, taking into account the theoretical errors. We note that, with none of the NV2+3 models considered are we able to reproduce the experimental strength of the first peak (see Fig.~\ref{fig:MFF_IIb_co}d), which we overpredict by ${\sim} 25-40\%$. For all the interactions considered, two-body currents give a correction of ${\sim} 25\%$ in the region of the first peak, and up to over ${\sim} 50\%$ in the tail. This is in line with the findings of Ref.~\cite{Donnelly1984}, which are based on phenomenological models. In that study, contributions from two-body meson exchange are found to be of the order of $7-15\%$ over most of the $q$ region. 

In $\besev$, the contribution from NLO two-body currents enhances the LO magnetic f.f. by a factor of ${\sim} 2$, this enhancement is even more pronounced when adding the N3LO currents, almost uniformly across the $0.5\lesssim q\lesssim 2.5$ fm$^{-1}$ region. This would make the $\besev$ magnetic f.f. a perfect candidate for validating models of two-body currents, if that experimental data were available.

\subsubsection{Lithium-8 and Boron-8}

There are no experimental data for the magnetic f.f.'s of the $A=8$ systems; however, from our calculations, we observe again a large contribution from two-body currents (Figs.~\ref{fig:MFF_IIb_co}, panels (f) and (g)). This is particularly striking in $^8$B, where two-body currents reduce the strength of the form factor for all $q$. It is also worth noting the difference in the structure imprinted on the calculations. $^8$B only has a single dominant peak at low $q$ generated by the $M_1$ multipole and a quenched $M_3$ contribution at higher $q$ with a double peak structure. Instead, $^8$Li exhibits a slightly two-peaked structure, where the second peak generated by the $M_3$ multipole is strongly suppressed. Should advancements in electron scattering on radioisotopes progress enough to measure these form factors, they would also provide interesting cases to better understand two-body currents. 

\subsubsection{$A=9$ nuclei}

For $A=9$ systems, there are data only for the $\benin$ magnetic f.f.~\cite{Goldemberg1963,Lapikas1975,Rand1966,Vanpraet1965}. The data present a plateau in the $0.3\lesssim q \lesssim 1$ fm$^{-1}$ region, and a peak around $q\sim 1.7$ fm$^{-1}$. In Fig.~\ref{fig:MFF_IIb_co}i, we compare our results with the data. Referring to this figure, we note that the NLO and N3LO two-body currents correct the strength of the plateau obtained with the LO current alone, leading to an improved agreement with the data. This finding confirms the hypothesis that two body currents are crucial to describe the plateau as conjectured in Ref.~\cite{Donnelly1984}. On the other hand, they (in particular, the NLO current) significantly contribute to the second peak, resulting in an overestimation of its maximum. Several phenomenological models  based on Woods-Saxon-type wave functions (see Ref.~\cite{Donnelly1984} and reference there in) are able to reproduce the second peak, though they underestimate the plateau. On the other hand, a projected Hartree-Fock calculation~\cite{Bouten:1969xmx} shows a result very similar to ours where the secondary peak is overestimated even when two-body currents are not included.

In Figs.~\ref{fig:MFF_IIb_co}i,~\ref{fig:MFF_IIb_co}j, and~\ref{fig:MFF_IIb_co}k, we show the results for the remaining $A=9$ nuclei.  While in $\bnin$ the two-body corrections are minimal over all the $q$ range, both $\linin$ and $\cnin$ receive large contributions form the NLO and N3LO two-body currents, and, therefore, these systems are good candidates for studies of higher order corrections in the currents. 

As per the multipole contributions displayed in Figs.~\ref{fig:MFF_multipole_IIb}e--\ref{fig:MFF_multipole_IIb}h, we note that all the $A=9$ present a similar structure. A first peak,  dominated by $M_1$, is found around $q\sim0.5$ fm$^{-1}$, and a second peak, dominated by $M_3$, appears at $q\sim1.5-1.8$ fm$^{-1}$. The latter fills in the diffraction minimum observed in $M_1$. The strength of the second peak is larger for $\benin$ and $\cnin$, while in $\linin$ and $\bnin$ the first peak is dominant. A similar feature is also observed for the mirror systems $^7$Li and $^7$Be, where the former is dominated by $M_1$ and the latter by $M_3$.

\begin{figure}
   \centering
   \includegraphics[width=\columnwidth]{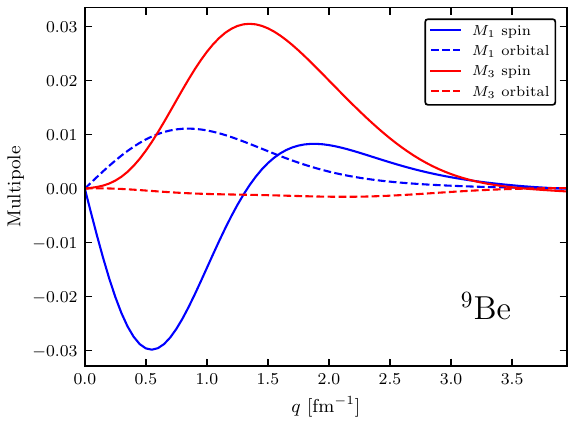}\\ 
   \includegraphics[width=\columnwidth]{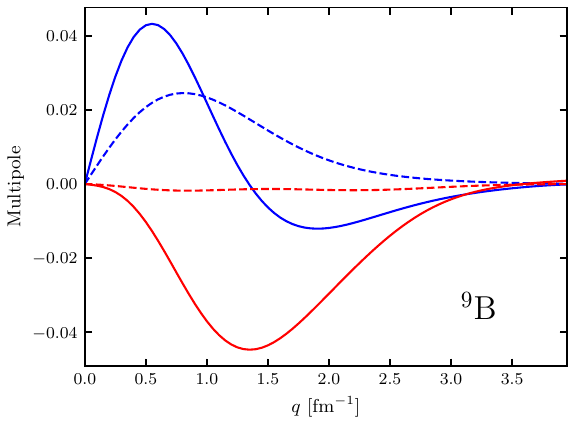} 
    \caption{Contributions of the spin (full) and orbit (dashed) terms to the
      $M_1$ (blue) and $M_3$ (red) multipoles at LO for the $^9$Be/$^9$B mirror system.}\label{fig:M_split}
\end{figure}

This pattern is generated by the interference between the spin and the orbital terms at LO (see Eq.~(\ref{eq:jlo})) in the $M_1$ multipole. In Fig.~\ref{fig:M_split}, as an example,  we compare the individual contributions to the $M_1$ and $M_3$ multipoles for the $^9$Be (unpaired neutron) and $^9$B (unpaired proton)  mirror systems. The spin terms (solid lines) for both the $M_1$ (blue curves) and $M_3$ (red curves) multipoles carry opposite sign in mirror nuclei due to the sign difference between the magnetic moments of the unpaired neutron and proton, respectively. At the same time, the $M_1$ multipole receives a substantial orbital contribution (dashed line) that has the same sign for both nuclei, though it is slightly suppressed when all protons are paired. Thus, the interference is destructive when only one neutron is unpaired, generating a suppression of the $M_1$; meanwhile, the case of an unpaired proton has constructive interference that generates an enhancement of the $M_1$ peak. Note that this effect is not present for the
$M_3$ multipole because the orbital term gives a negligible contribution and thus there are no interference effects. Similar findings have been observed for the other pairs mirror nuclei,  $^7$Li-$^7$Be and $^9$Li-$^9$C.

\subsubsection{Boron-10}

 The experimental data for the $\bten$ magnetic f.f. are provided for $0.5\lesssim q \lesssim 2.0$ fm$^{-1}$ region~\cite{Goldemberg1963,Goldemberg1965,Rand1966,Vanpraet1965,Lapikas1978}. We did not include the data of Refs.~\cite{Goldemberg1963,Goldemberg1965} in the plots because of the strong disagreement with the more recent experimental results reported in Ref.~\cite{Donnelly1984}.
The VMC calculation has a very nice agreement with the data in the  region of the main peak (see Fig.~\ref{fig:MFF_IIb_co}l). Contributions coming from two-body currents are very small because they are purely IS. As shown in Fig.~\ref{fig:MFF_multipole_IIb}i, the main peak is dominated by the $M_1$ multiple, while the secondary peak in the $q\sim2.5-3$ fm$^{-1}$ region is generated by the $M_5$ multipole that also contributes to the filling in of the $M_1$'s diffraction minimum. Note that this secondary peak  was hypothesized in Ref.~\cite{Donnelly1984}, but it was not found in shell-model calculations performed in that reference. The $M_3$ peak that dominates all the other nuclei in the  $q\sim 1.5$ fm$^{-1}$ region is barely visible, as in Ref.~\cite{Donnelly1984}.

\section{Conclusions}
\label{sec:conclusion}
In this work we presented QMC calculations of magnetic moments and magnetic form factors of nuclei
for $A\leq10$ based on chiral effective field theory. For the calculation we used the Norfolk two- and three- body interaction and consistent
one- and two-body current operators.

The computation of magnetic moments in light nuclei displayed the important role of two-body contributions. Consistent with previous QMC studies~\cite{Pastore:2012rp,Martin:2023dhl}, two-body currents can make up a significant fraction (${\approx}$30\% -- 40\%) of the total magnetic moments in light systems. The analysis of two-body currents in this work was supplemented by the computation of two-body magnetic densities in a microscopic framework. These densities not only corroborated the understanding of the two-body contributions on the basis of the dominant cluster structures in nuclear systems, but also further illuminated our understanding of how the short-range dynamics impact the system. As for transition densities associated with $\beta$-decay~\cite{King:2020wmp}, magnetic densities display a striking universal behavior at short distances. The long-range structure, however, is sensitive to the details of pair formation inside the nucleus which is driven by the tensor component of the nuclear force. Finally, these densities displayed how changing the cutoff in the Norfolk model influences the short-range dynamics associated with this observable. 

Consistent with the studies of Refs.~\cite{Gnech:2022vwr,Martin:2023dhl}, we find that fitting the LECs appearing at N3LO in the electromagnetic current can lead to unnaturally large contributions compared to the expectation from naive dimensional analysis. In the Norfolk model, this is driven by a rather large contribution from the short-range N3LO(NM) diagram. This enhancement spoils the order-by-order convergence in these calculation in the Norfolk model, as is also the case for the other chiral potentials appearing in the previous QMC studies. 

We investigated the impact of different regulators in the $\chi$EFT approach by looking at different Norfolk models. As in the calculations performed in Ref.~\cite{Pal:2023gll}, a regulator dependence at NLO also appears in our calculations. Combined with the lack of convergence in the power counting, these results merit further scrutiny to better understand the physics responsible for these phenomena. A detailed investigation of the regulator dependence in long-range terms could illuminate whether this behavior is due to power corrections from the choice of cutoff or to a lack of renormalization in the framework. 

To validate away from zero momentum transfer, we performed the calculation of magnetic form factors. In $A=3$ systems, our results represent a benchmark with Ref.~\cite{Gnech:2022vwr} and we find good agreement with the previous evaluation and the data. We additionally performed the first microscopic calculations of magnetic form factors in $6\,\leq A\,\leq 10$ nuclei using a $\chi$EFT model, which we presented in a companion publication~\cite{prl}. We observed good qualitative agreement with the data in most cases, reproducing the strength and position of peaks in the measured form factor. We also observed for the first time that, in the mirror nuclei, there is an inversion of the strength of the $M_1$ and $M_3$ multipoles. Here, we supplemented the discussion of Ref.~\cite{prl} by demonstrating that this inversion is due to an interference between the LO orbital and spin-magnetization contributions to the $M_1$ multipole. As no experiments or previous calculations exist for these mirror systems, this represents a prediction of this effect. 

Finally, our calculations of the magnetic form factor also included an analysis of the contributions from two-body currents. As for the magnetic moments, two-body currents make a sizeable contribution to the form factors. Over the range of momenta studied, two-body currents can be as large as ${\approx}$20\% -- 50\%.

Because of the interesting structural features and the insights provided on two-body currents, it is our view that a renewed interest in studying elastic magnetic form factors would spark further improvements in the $\chi$EFT framework. Until recently, it was only possible to measure these data for stable systems; however, novel techniques were recently used to perform the first electron scattering measurement for a radioisotope~\cite{Tsukada:2023}. In the future, a synergy between experimentalists and theorists could identify which systems represent feasible and interesting candidates for such a study. This type of effort, aided by the current and future capabilities of state-of-the-art rare isotope beam facilities, will only serve to enhance our understanding of models of nuclear structure, interactions, and currents.

\acknowledgments 

The authors would like to thank Josh Martin, Emanuele Mereghetti, and Ingo Tews for useful discussions at various stages in the preparation of this manuscript.

 This work is supported by the US Department of Energy under Contracts No. DE-SC0021027 (G.~C.-W., G.~B.~K. and S.~P.), DE-AC02-06CH11357 (R.B.W.), DE-AC05-06OR23177 (A.G. and R.S.), a 2021 Early Career Award number DE-SC0022002 (M.~P.), the FRIB Theory Alliance award DE-SC0013617 (M.~P.), and the NUCLEI SciDAC program (S.P., M.P., and R.B.W.). G.~B.~K. would like to acknowledge support from the U.S. DOE NNSA Stewardship Science Graduate Fellowship under Cooperative Agreement DE-NA0003960. G.~C.-W. acknowledges support from the NSF Graduate Research Fellowship Program under Grant No. DGE-213989. We thank the Nuclear Theory for New Physics Topical Collaboration, supported by the U.S.~Department of Energy under contract DE-SC0023663, for fostering dynamic collaborations.
 A.G. acknowledges the direct support of Nuclear Theory for New Physics Topical collaboration.

The many-body calculations were performed on the parallel computers of the Laboratory Computing Resource Center, Argonne National Laboratory, the computers of the Argonne Leadership Computing Facility (ALCF) via the INCITE grant ``Ab-initio nuclear structure and nuclear reactions'', the 2019/2020 ALCC grant ``Low Energy Neutrino-Nucleus interactions'' for the project NNInteractions, the 2020/2021 ALCC grant ``Chiral Nuclear Interactions from Nuclei to Nucleonic Matter'' for the project ChiralNuc, the 2021/2022 ALCC grant ``Quantum Monte Carlo Calculations of Nuclei up to $^{16}{\rm O}$ and Neutron Matter" for the project \mbox{QMCNuc}, and by the National Energy Research
Scientific Computing Center, a DOE Office of Science User Facility
supported by the Office of Science of the U.S. Department of Energy
under Contract No. DE-AC02-05CH11231 using NERSC award
NP-ERCAP0027147.

\appendix

\section{Electromagnetic currents in configuration space }
\label{app:currr}

In this appendix, we collect the $r$-space expressions of the electromagnetic current operators used in this work, which are currently dispersed across various references~\cite{Pastore:2008ui,Pastore:2009is,Pastore:2011ip,Piarulli:2012bn,Schiavilla:2018udt,Gnech:2022vwr}. The electromagnetic current operators are written as an expansion in a generic low momentum scale $Q$. 

The LO ($Q^{-2}$) one-body current operator, displayed in panel (a) of Fig.~\ref{fig:currn2lo}, is given by
\begin{equation}
\label{eq:jlo}
{\bf j}^{\rm LO}({\bf q})=\frac{\epsilon_i(q_{\mu}^2) }{2\, m}\,
 \left[{\bf p}_i\,\, ,\, \, {\rm e}^{i{\bf q}\cdot{\bf r}_i}  \right]_+ 
+i\,\frac{\mu_{i}(q_{\mu}^2) }{2\, m} \,\, {\rm e}^{i{\bf q}\cdot{\bf r}_i}\,\, {\bm \sigma}_i\times {\bf q }\ ,
 \end{equation}
where ${\bf p}_i\,$=$\, -i\, {\bm \nabla}_i$, ${\bf r}_i$ is the single-nucleon coordinate,  and $m$ is the average nucleon mass. The brackets $[ \ldots,\ldots]_{+}$ indicate the anticommutator and the operators $\epsilon_i$ and $\mu_i$ are isospin dependent operators defined as
\begin{eqnarray}
\epsilon_i&=& \frac{1}{2}\left(G^S_E(q_{\mu}^2)+G^V_E(q_{\mu}^2)\tau_{i,z}\right) \, , \\
\mu_i &=& \frac{1}{2}\left(G^S_M(q_{\mu}^2)+G^V_M(q_{\mu}^2)\tau_{i,z}\right) \ .
\end{eqnarray}
The isoscalar/isovector combinations of the proton (p) and neutron (n) electric (E) and magnetic (M) form factors are defined as
\begin{equation}
G^{S/V}_{E/M} = G^p_{E/M}(q_{\mu}^2) \pm G^n_{E/M}(q_{\mu}^2) \, ,  
\end{equation}
which depend on $q_{\mu} = q^2 - \omega^2$, with $\omega$ being the energy transfer. Though the form factors, which appear in all order of the the current, themselves have a power series expansion in $Q$, this is typically neglected and they are treated as multiplicative factors. The form factors in this work are not taken from chiral perturbation theory, but rather from experimental fits of elastic scattering on the proton and deuteron. We adopt the same form factors as in Ref.~\cite{Gnech:2022vwr}, which are given by
\begin{eqnarray*}
G_E^p(q_{\mu}^2) &=& G_D(q_{\mu}^2)\, ,\\
G_E^n(q_{\mu}^2) &=& -\mu^n \chi G_D(q_{\mu}^2) (1 + 4\chi)^{-1} \, ,\\
G_M^p(q_{\mu}^2) &=& \mu_p G_D(q_{\mu}^2)\, , \\
G_M^n(q_{\mu}^2) &=& \mu_n G_D(q_{\mu}^2)\, ,
\end{eqnarray*}
where $\mu_p$ and $\mu_n$ are the proton and neutron magnetic moments, respectively, $\chi \equiv q_{\mu}^2/(4m_N^2)$, and the dipole form factor is defined as
\begin{equation}
G_D(q_{\mu}^2) = (1 + q_{\mu}^2/\Lambda^2)^{-2}\, ,    
\end{equation}
for $\Lambda = 0.83$ GeV.

At NLO ($Q^{-1}$), the currents receive a purely IV, long-range OPE contribution--see panels (b) and (c) of Fig.~\ref{fig:currn2lo}--given by 
\begin{eqnarray}
{\bf j}^{\rm NLO}_{\rm OPE}({\bf q})&=& G^V_E(q_{\mu}^2) \times \\
&& \bigg[ {\rm e}^{i\,{\bf q}\cdot {\bf r}_i}\,
 \left({\bm \tau}_i \times {\bm \tau}_j\right)_z
\, I_0^\pi(x_{ij}) \, {\bm \sigma}_i \,\, {\bm \sigma}_j \cdot \hat{\bf r}_{ij}+\! (i \rightleftharpoons j)  \nonumber \\
&&+\, {\rm e}^{i\,{\bf q}\cdot {\bf R}_{ij}}\, \left({\bm \tau}_i \times {\bm \tau}_j\right)_z\, 
 \,{\bm \sigma}_i \cdot \left({\bm \nabla}^x_{ij} + i\, \frac{\bf q}{2\, m_\pi}\right) \nonumber\\
&&\times {\bm \sigma}_j \cdot \left({\bm \nabla}_{ij}^x -i\, \frac{\bf q}{2\, m_\pi}\right)
 {\bm \nabla}^x_{ij}\, L_0^\pi({\bm x}_{ij},{\bf q})\bigg] \ ,
\label{eq:nlor2}
\end{eqnarray}
where $\bfx_{ij}=\bfr_{i}-\bfr_{j}$ and $\bfR_{ij}=(\bfr_i+\bfr_j)/2$ are the relative and center of mass coordinates of nucleons $i$ and $j$, respectively, while $m_\pi$ denotes the pion mass. The gradients ${\bm \nabla}^x_{ij}$ are taken with respect to the adimensional variables ${\bm x}_{ij}=m_{\pi}\bfr_{ij}$.  The correlation functions
$I_0^\pi (x)$ and $L_0^\pi({\bm x},{\bf q})$, where we dropped the indices $ij$ for brevity, are defined as
\begin{eqnarray}
\label{eq:e12}
I_0^\pi (x)&=& -\frac{g^2_A}{16\, \pi}\, \frac{m_\pi^2}{f_\pi^2}\, (1+x)\, \frac{e^{-x}}{x^2} \ ,\\
\label{eq:e11} \nonumber
L_0^{\pi}({\bm x},{\bf q}) &=&\frac{g^2_A}{16\, \pi}\, \frac{m_\pi^2}{f_\pi^2} \\
&&\int_{-1/2}^{1/2} dz\,\,
 {\rm e}^{-i\left(\frac{z}{m_\pi}\right)\bfq \cdot {\bm x}} \,\, \frac{e^{-x \, L(z,q)}}{ L(z,q)} \ , 
\end{eqnarray}
with
\begin{equation}
 L(z,q)=\sqrt{1+\frac{q^2}{4\, m_\pi^2}\left(1-4\, z^2\right)} \ .
 \end{equation}
In the preceding expressions, $g_A=1.29$ is the nucleon axial coupling constant corrected to account for the Goldberger-Treiman discrepancy~\cite{Arndt:1994bu,Stoks:1992ja}.
 
The N2LO ($Q^{0}$) current involves a one-body relativistic correction (RC), shown in panel (d) of Fig.~\ref{fig:currn2lo}, and an IV term of one-pion range involving the excitation of an intermediate $\Delta$-isobar, displayed in panel (e) of Fig.~\ref{fig:currn2lo}. Their expressions are 
\begin{widetext}
\begin{eqnarray}
\label{eq:j1rcc}
{\bf j}^{\rm N2LO}_{\rm RC}({\bf q})&=&-\frac{\epsilon_i(q_{\mu}^2)}{16 \, m^3}
 \left[2 \left( p_i^2 +\frac{q^2}{4} \right)  \big( 2\, {\bf p}_i
+i\, {\bm \sigma}_i\times {\bf q } \big) 
+ {\bf p}_i\cdot {\bf q}\, \left({\bf q} +2\, i\, {\bm \sigma}_i\times {\bf p }_i \right)
 \,\, ,\,\, 
 {\rm e}^{i {\bf q}\cdot {\bf r}_i}\right]_{+}\nonumber \\
 && \hspace{-0.2cm} - i\, \frac{\mu_i(q_{\mu}^2)-\epsilon_i(q_{\mu}^2)}{16 \, m^3}
  \bigg[ {\bf p}_i\cdot {\bf q} \big( 4\, {\bm \sigma}_i\times {\bf p}_i-i\, {\bf q}\big) \nonumber \\
 &&\left. - \left(  2\, i\, {\bf p}_i -{\bm \sigma}_i\times {\bf q} \right)\frac{q^2}{2}  +2\, \left({\bf p}_i\times {\bf q}\right)
 \, {\bm \sigma}_i\cdot {\bf p}_i
 \,\, , \,\, {\rm e}^{i {\bf q}\cdot {\bf r}_i}\right]_{+}  \ , \\
{\bf j}^{\rm N2LO}_{\Delta}({\bf q})&=&-iG_{\gamma N \Delta}(q_{mu}^2)\bigg\{ {\rm e}^{i {\bf q}\cdot {\bf r}_i} \, \tau_{j,z}
\left[ I^{\Delta}_1(x_{ij}) \, {\bm \sigma}_j
+I_{2}^{\Delta}(x_{ij}) \,\,
 {\bm \sigma}_j\cdot \hat{\bf r}_{ij} \,\, \hat{\bf r}_{ij}\right] \nonumber\\
&&- \frac{1}{4}\, {\rm e}^{i\, {\bf q}\cdot {\bf r}_i}\, \left({\bm \tau}_i\times{\bm \tau}_j\right)_z \left[
I^{\Delta}_1 (x_{ij})\, {\bm \sigma}_i \times {\bm \sigma}_j
+I^\Delta_2(x_{ij})\, {\bm \sigma}_j\cdot \hat{\bf r}_{ij}\,\, {\bm \sigma}_i\times \hat{\bf r}_{ij}\right]\bigg\} \times\frac{{\bf q}}{m_\pi} \nonumber \\
&& +(i\rightleftharpoons j )\ , \label{eq:jdn2lo}
\end{eqnarray}
\end{widetext}
with the correlation functions $I^\Delta_k(x)$ defined as follows,
\begin{eqnarray}
\label{eq:e13}
I^\Delta_1(x)&=&- \left(\frac{g_A\,h_A}{18\, \pi}\,
\frac{1}{2\, m}\, \frac{m_\pi^2}{m_{\Delta N}}
 \frac{m_\pi^2}{f_\pi^2}\right) \nonumber \\
 &&\hspace{1cm} \times (1+x)\, \frac{e^{-x}}{x^3} \ , \\
\label{eq:e14}
I^\Delta_2(x)&=&\left(\frac{g_A\,h_A}{18\, \pi}\,
\frac{1}{2\, m}\, \frac{m_\pi^2}{m_{\Delta N}}
 \frac{m_\pi^2}{f_\pi^2}\right) \nonumber \\
&&\hspace{1cm} \times (3+3\,x+x^2)\, \frac{e^{-x}}{x^3} \ ,
\end{eqnarray}
where $m_{\Delta N}=0.2931$ GeV is the delta-nucleon mass difference and $h_A=2.73$ is the $\pi N\Delta$ coupling constant.  The $N$-to-$\Delta$ transition form factor is defined as,
\begin{equation}
G_{\gamma N \Delta}(q_{mu}^2) = \frac{\mu_{\gamma N \Delta}}{(1 + q_{\mu}^2/\Lambda_{\Delta,1}^2)\sqrt{1+q_{\mu}^2/\Lambda_{\Delta,2}^2}}\, ,
\end{equation}
where the transition magnetic moment $\mu_{\gamma N \Delta} = 3 \mu_N$ is taken from an analysis of $\gamma N$ data in the $\Delta$ resonance region~\cite{Carlson:1985mm} and we adopt $\Lambda_{\Delta,1}=0.84$ GeV and $\Lambda_{\Delta,2}=1.2$ GeV from the same analysis.

At N3LO ($Q^1$), the long-range current operator consists of an isocalar (IS) and an isovector (IV) OPE terms generated by the diagram illustrated in panel (f) of Fig.~\ref{fig:currn3lo}. The N3LO IS current of one-pion range is proportional to the $d_9^\prime$ LEC, and it reads
\begin{widetext}
\begin{eqnarray}
{\bf j}^{\rm N3LO}_{\rm IS~OPE}({\bf q})&=&-i\,{\rm e}^{i {\bf q}\cdot {\bf r}_i} G_{\gamma \pi \rho}(q_{\mu}^2)\,  {\bm \tau}_i\cdot{\bm \tau}_j
\left[ I^S_1(x_{ij}) \, {\bm \sigma}_j
+I^S_2(x_{ij}) \,\,
 {\bm \sigma}_j\cdot \hat{\bf r}_{ij} \,\, \hat{\bf r}_{ij}\right]\times 
 \frac{{\bf q}}{m_\pi}+(i\rightleftharpoons j )\ ,
 \label{eq:n3lo.is} 
\end{eqnarray}
\end{widetext}
where the correlation functions, namely $I^S_1(x)$ and $I^S_2(x)$,  may be obtained from the following rescaling of the N2LO($\Delta$) correlation functions of Eqs.~(\ref{eq:e13})-(\ref{eq:e14}) 
\begin{equation}
\left(\frac{g_A\,h_A}{18\, \pi}\,
\frac{1}{2\, m}\, \frac{m_\pi^2}{m_{\Delta N}}
 \frac{m_\pi^2}{f_\pi^2}\right) 
 \longrightarrow \left(\frac{g_A}{16\, \pi}\,
 \frac{m_\pi^2}{f_\pi^2}\, m_\pi^2\, d_9^\prime \right) \, , \nonumber
\end{equation}
and the transition form factor is defined as,
\begin{equation}
G_{\gamma \pi \rho}(q_{\mu}^2) = \frac{1}{1+q_{\mu}^2/m_{\omega}^2}\, ,
\end{equation}
where $m_{\omega}$ is the $\omega$-meson mass. 

The N3LO IV current of one-pion range is composed by two terms proportional to the $d_8^\prime$ and $d_{21}^\prime$ LECs, respectively:
\begin{equation}
{\bf j}^{\rm N3LO}_{\rm IV~OPE}({\bf q}) = {\bf j}^{{\rm N3LO}, \,d_8^\prime}_{\rm IV~OPE}({\bf q}) + {\bf j}^{{\rm N3LO}, \,d_{21}^\prime}_{\rm IV~OPE}({\bf q})\, \label{eq:n3lo.iv},
\end{equation}
where 
\begin{widetext}
\begin{eqnarray}
 {\bf j}^{{\rm N3LO}, \,d_8^\prime}_{\rm IV~OPE}({\bf q})&=&-i\,{\rm e}^{i {\bf q}\cdot {\bf r}_i}\frac{G_{\gamma N \Delta}(q_{\mu}^2)}{\mu_{\gamma N \Delta}} \, \tau_{j,z}
\left[ I^V_1(x_{ij}) \, {\bm \sigma}_j
+I_{2}^V(x_{ij}) \,\,
 {\bm \sigma}_j\cdot \hat{\bf r}_{ij} \,\, \hat{\bf r}_{ij}\right]\times 
 \frac{{\bf q}}{m_\pi} +(i\rightleftharpoons j )\, , \label{eq:n3lo.iv.d8}\\
 {\bf j}^{{\rm N3LO}, \,d_{21}^\prime}_{\rm IV~OPE}({\bf q})&=& \frac{i}{4}\, {\rm e}^{i\, {\bf q}\cdot {\bf r}_i}\frac{G_{\gamma N \Delta}(q_{\mu}^2)}{\mu_{\gamma N \Delta}} \, \left({\bm \tau}_i\times{\bm \tau}_j\right)_z \left[
\widetilde{I}^{V}_1 (x_{ij})\, {\bm \sigma}_i \times {\bm \sigma}_j
+\widetilde{I}^V_2(x_{ij})\, {\bm \sigma}_j\cdot \hat{\bf r}_{ij}\,\, {\bm \sigma}_i\times \hat{\bf r}_{ij}\right] \times\frac{{\bf q}}{m_\pi} \nonumber \\
&&\hspace{8cm}+(i\rightleftharpoons j )\ . \label{eq:n3lo.iv.d21} 
\end{eqnarray}
\end{widetext}
In the expressions above, the IV correlation functions $I^V_1(x)$ and $I^V_2(x)$ are obtained from the correlation functions of Eqs.~(\ref{eq:e13})-(\ref{eq:e14}) by substituting,
\begin{equation}\label{eq:d8}
\left(\frac{g_A\,h_A}{18\, \pi}\,
\frac{1}{2\, m}\, \frac{m_\pi^2}{m_{\Delta N}}
 \frac{m_\pi^2}{f_\pi^2}\right) 
 \longrightarrow \left(\frac{g_A}{16\, \pi}\,
 \frac{m_\pi^2}{f_\pi^2}\, m_\pi^2\, d_8^\prime \right) \ .
\end{equation}
while the correlation functions $\widetilde{I}^V_1(x)$ and $\widetilde{I}^V_2(x)$ can be obtained by making the substitution, 
\begin{equation}\label{eq:d21}
\left(\frac{g_A\,h_A}{18\, \pi}\,
\frac{1}{2\, m}\, \frac{m_\pi^2}{m_{\Delta N}}
 \frac{m_\pi^2}{f_\pi^2}\right) 
 \longrightarrow \left(\frac{g_A}{16\, \pi}\,
 \frac{m_\pi^2}{f_\pi^2}\, m_\pi^2\, d_{21}^\prime \right) \ .
\end{equation}

The N3LO TPE loop term coming from the contributions shown in panels (g)-(k) of Fig.~\ref{fig:currn3lo} has the following form
\begin{widetext}
\begin{eqnarray}
{\bf j}^{{\rm N3LO}}_{\rm LOOP}({\bf q})&=&
i\,\tau_{j,z}\,\, {\rm e}^{i{\bf q}\cdot {\bf R}_{ij}}G^V_E(q_{\mu}^2)
\left\{ \left(
\left[ F^{(0)}_0(\lambda_{ij})  + F^{(1)}_2(\lambda_{ij})\right]{\bm \sigma}_i 
+ F^{(2)}_2(\lambda_{ij})\,   {\bm \sigma}_i\cdot\hat{\bf r}_{ij} \,\, \hat{\bf r}_{ij}
 \right) \times \frac{{\bf q}}{2\, m_\pi}  \right. \nonumber\\
&&\left. \hspace{-0.2cm}-\frac{1}{2} ({\bm \tau}_i\times {\bm \tau}_j)_z\,\,  {\rm e}^{i{\bf q}\cdot {\bf R}_{ij}} \,\hat{\bf r}_{ij}\,
\left[ \lambda_{ij}\, \frac{\tilde{v}_{\,2\pi}(\lambda_{ij})}{2\,m_\pi} \right]\right\}
+ (i \rightleftharpoons j) \ . \label{eq:jloop}
\end{eqnarray}
\end{widetext}

A detailed derivation of the loop functions $F^{(n)}(\lambda_{ij})$, where $\lambda_{ij}=2m_{\pi}r_{ij}$, can be found in  Ref.~\cite{Schiavilla:2018udt}. Here, we limit ourselves to listing their expressions,
\begin{widetext}
\begin{eqnarray*}
F_0^{(0)}(\lambda) &=& \frac{g_A^4}{256\,\pi^3}\,
\frac{\left(2\, m_\pi\right)^4}{f_{\pi}^4}\bigg\{
  \frac{{\rm e}^{-\lambda}}{\lambda} \left[ 1+\frac{1}{2}\int_0^1dz\, \frac{1-{\rm e}^{-\lambda(\alpha_z-1)}}{\alpha_z^2-1} \right]
-\left[\left(\frac{1}{g_A^2} +1\right)\frac{1}{\lambda} -\frac{1}{2} \right]  E^{(-)}_1(\lambda) \nonumber\\
&& \hspace{7cm} -\left[ \left( \frac{1}{g_A^2}+1\right)\frac{1}{\lambda}+\frac{1}{2}\right]
 E^{(+)}_1(\lambda)  \bigg\}\ , \\
F_1^{(2)}(\lambda) &=& \frac{g_A^4}{256\,\pi^3}\,\frac{\left(2\, m_\pi\right)^4}{f_{\pi}^4} \bigg\{
  {\rm e}^{-\lambda} \left(\frac{1}{\lambda^3}+\frac{1}{\lambda^2}\right) \left[1+\frac{1}{2}\int_0^1dz\, \frac{1-{\rm e}^{-\lambda(\alpha_z-1)}}{\alpha_z^2-1} \right] \nonumber\\
  &&\hspace{-0.25cm}- \left[ \left( \frac{1}{g_A^2}+2\right)
 \left(\frac{1}{\lambda^3}+\frac{1}{\lambda^2}\right) 
-\frac{1}{2\,\lambda}\right] \! E_1^{(-)}(\lambda)
- \left[ \left( \frac{1}{g_A^2}+2\right) 
\left(\frac{1}{\lambda^3}-\frac{1}{\lambda^2}\right)-\frac{1}{2\,\lambda}\right] 
\! E_1^{(+)}(\lambda)\nonumber\\
&&\hspace{-0.25cm}+ \left(\frac{1}{g_A^2} -2\right)\,\frac{1}{\lambda^3}\, E_1^{(0)}(\lambda)
-\int_0^1 dz \left(\frac{4}{\lambda^3}+\frac{1}{\alpha_z+1}\, \frac{1}{2\, \lambda^2} \right)
 {\rm e}^{-\lambda\alpha_z}\bigg\}\ , \\
F_2^{(2)}(\lambda) &=&  -\frac{g_A^4}{256\,\pi^3}\,\frac{\left(2\, m_\pi\right)^4}{f_{\pi}^4} \bigg\{
  {\rm e}^{-\lambda} \left(\frac{3}{\lambda^3}+\frac{3}{\lambda^2}+\frac{1}{\lambda}\right) \left[1+\frac{1}{2}\int_0^1dz\, \frac{1-{\rm e}^{-\lambda(\alpha_z-1)}}{\alpha_z^2-1} \right] \nonumber\\
&&\hspace{3cm}- \left[ \left( \frac{1}{g_A^2}+2\right)
 \left(\frac{3}{\lambda^3}+\frac{3}{\lambda^2}+\frac{1}{\lambda}\right) 
-\frac{1}{2}\left(1+ \frac{1}{\lambda}\right)\right] \! E_1^{(-)}(\lambda) \nonumber\\
&&\hspace{3cm}-\left[ \left( \frac{1}{g_A^2}+2\right)
 \left(\frac{3}{\lambda^3}-\frac{3}{\lambda^2}+\frac{1}{\lambda}\right) 
+\frac{1}{2}\left(1- \frac{1}{\lambda}\right)\right]   \! E_1^{(+)}(\lambda)
+ \left(\frac{1}{g_A^2} -2\right)\frac{3}{\lambda^3} \, E_1^{(0)}(\lambda) \nonumber\\
&&\hspace{3cm}-\int_0^1 dz \left[\frac{16}{\lambda^3}
+\left( 4\,\alpha_z+\frac{3}{2}\, \frac{1}{\alpha_z+1}\right) 
\frac{1}{\lambda^2}\right]
{\rm e}^{-\lambda\alpha_z} \bigg\}\ ,
\end{eqnarray*}
\end{widetext}
where the following definitions have been employed,
\begin{eqnarray}
\alpha_z &=& (1-z^2)^{-1/2} \, , \\
E_1(x) &=& \int_x^{\infty} dt \frac{e^{-t}}{t} \, , \\
E_1^{(\pm)}(\lambda) &=& \frac{e^{\pm \lambda}}{2}\int_0^1 dz E_1(\lambda\alpha_z + \lambda) \, \\
E_1^{(0)}(\lambda) &=& \int_0^1 dz E_1(\lambda \alpha_z)
\end{eqnarray}

The N3LO LOOP two-body currents satisfy the continuity equation with the NV2 potential at N2LO in the limit of momentum transfer going to zero. This is reflected by the presence of the $\tilde{v}_{2\pi}(\lambda_{ij}) = v_{2\pi}(\lambda_{ij})\bfta_i\cdot\bfta_j$ term in Eq.~(\ref{eq:jloop}).  The $v_{2\pi}(\lambda_{ij})$ term indicates contributions to the NV2 potential coming from TPE diagrams. In the present study, it accounts for TPE terms at both NLO and N2LO, therefore it includes two-pion loops with and without $\Delta$-isobar intermediate states (see Ref.~\cite{Schiavilla:2018udt} for more details).

The last contribution at N3LO is displayed in panel (l) of Fig.~\ref{fig:currn3lo}. This short-range contribution is divided into i) a term obtained via minimal substitution (MIN) in the NV2 contact operators, ii) a non-minimal term (NM), and iii) a term used to regularize the N3LO loop contribution (CT). 

The N3LO(MIN) contact term has IS and IV components: 
\begin{equation}
    {\bf j}^{\rm N3LO}_{\rm MIN}({\bf q})={\bf j}^{\rm N3LO}_{\rm MIN\, IS}({\bf q})+{\bf j}^{\rm N3LO}_{\rm MIN\, IV}({\bf q})\, \label{eq:jmin},
\end{equation}
respectively given by
\begin{widetext}
\begin{eqnarray}
{\bf j}^{\rm N3LO}_{\rm MIN\,IS}({\bf q})&=&
-\frac{1}{8} m_\pi^4\, C_5 \,{\rm e}^{i{\bf q}\cdot{\bf R}_{ij}}G^S_E(q_{\mu}^2) \Big[  i\, C^{(0)}_{R_{\rm S}}(z_{ij})\left({\bm \sigma}_i+{\bm \sigma}_j\right) \times \frac{\bf q}{m_\pi}\Big] \ ,\label{eq:jmin.is} \\
{\bf j}^{\rm N3LO}_{\rm MIN\, IV}({\bf q})&=&\frac{1 }{8}\,  \,
\left({\bm \tau}_i\times{\bm \tau}_j\right)_z\,\,
{\rm e}^{i{\bf q}\cdot{\bf R}_{ij}} G^V_E(q_{\mu}^2)\
C^{(1)}_{R_{\rm S}}(z_{ij}) 
\Big[ \, m_\pi^4\left( C_2+3\, C_4+C_7\right) \hat {\bf r}_{ij} \nonumber \\
&& + m_\pi^4\left( C_2-C_4-C_7\right)  \,\hat {\bf r}_{ij} \, \,{\bm \sigma}_i\cdot{\bm \sigma}_j + \, m_\pi^4\, C_7 \left( {\bm \sigma}_i\cdot \hat{\bf r}_{ij}\,\, {\bm \sigma}_j
+ {\bm \sigma}_j\cdot \hat{\bf r}_{ij}\,\, {\bm \sigma}_i\right) \Big] \nonumber \\
&& -\frac{1}{8} m_\pi^4\, C_5 \,{\rm e}^{i{\bf q}\cdot{\bf R}_{ij}} \Big[ \left(\tau_{i,z}-\tau_{j,z}\right)\, 
C^{(1)}_{R_{\rm S}}(z_{ij}) 
 \left({\bm \sigma}_i+{\bm \sigma}_j\right)
\times \hat{\bf r}_{ij}  \Big] \ .\label{eq:jmin.iv} 
\end{eqnarray}
\end{widetext}
In the equations above
\begin{eqnarray}
C^{(0)}_{R_{\rm S}} &=& \frac{e^{-z^2}}{\pi^{3/2}(m_{\pi}R_S)^3} \, ,\\
C^{(1)}_{R_{\rm S}} &=& \frac{1}{m_{\pi}R_S} \frac{dC^{(0)}_{R_S}}{dz}\, , 
\end{eqnarray}
where $z_{ij}=r_{ij}/R_{\rm S}$.
The $C_i$ LECs  in Eqs.~(\ref{eq:jmin.is})-(\ref{eq:jmin.iv}) are the same that appear in the contact terms of the NV2 potential at N2LO.

The N3LO(NM) contact term is also broken into IS and IV components as
\begin{equation}
    {\bf j}^{\rm N3LO}_{\rm NM}({\bf q})={\bf j}^{\rm N3LO}_{\rm NM\, IS}({\bf q})+{\bf j}^{\rm N3LO}_{\rm NM\, IV}({\bf q})\, \label{eq:jnm},
\end{equation}
where
\begin{widetext}
\begin{eqnarray}
 {\bf j}^{\rm N3LO}_{\rm NM\,IS}({\bf q})&=& - i \, {\rm e}^{i{\bf q}\cdot{\bf R}_{ij}}\, 
C^{(0)}_{R_{\rm S}}(z_{ij}) \,  m_\pi^4\, G^S_E(q_{\mu}^2)\, C_{15}^\prime 
\left({\bm \sigma}_i+{\bm \sigma}_j\right) \times \frac{{\bf q}}{m_\pi}  \, ,\label{eq:jnm.is}\\
 {\bf j}^{\rm N3LO}_{\rm NM\, IV}({\bf q})&=& - i \, {\rm e}^{i{\bf q}\cdot{\bf R}_{ij}}\, 
C^{(0)}_{R_{\rm S}}(z_{ij}) \,  m_\pi^4\, G^V_E(q_{\mu}^2)\, C_{16}^\prime\, (\tau_{i,z} - \tau_{j,z})  \left({\bm \sigma}_i-{\bm \sigma}_j\right)
 \times \frac{{\bf q}}{m_\pi}  \, .\label{eq:jnm.iv}
\end{eqnarray}
\end{widetext}
The last N3LO contact term, originating from  the regularization scheme in configuration space adopted for the TPE currents~\cite{Schiavilla:2018udt}, reads: 
\begin{widetext}
\begin{eqnarray}
{\bf j}^{{\rm N3LO}}_{\rm CT}({\bf q})&=&
iG^V_E(q_{\mu}^2)\,\tau_{j,z}\,\, {\rm e}^{i{\bf q}\cdot {\bf R}_{ij}}
\,\, F^{(0)}_0(z_{ij};\infty)\,\, {\bm \sigma}_i  \times \frac{{\bf q}}{2\, m_\pi} 
+ (i \rightleftharpoons j) \,   , \label{eq:jloopct}
\end{eqnarray}
\end{widetext}

For the unknown LECs of non-minimal nature, namely, $d_8^\prime$, $d_9^\prime$, $d_{21}^\prime$, $C_{15}^\prime$ and $C_{16}^\prime$, we take the values obtained in Ref.~\cite{Gnech:2022vwr} that use the deuteron and the three-nucleon systems' magnetic moments, as well as deuteron photo-disintegration data at backward angles to constrain them. These values are provided in Table~\ref{tab:fitNV}.

\begin{table}
  \begin{center}
    \begin{tabular}{lcccccc}
      \hline\hline
        Class  & $C_{16}^\prime m_\pi^4$ & $d_8^\prime m_\pi^2$ & $d_{21}^\prime m_\pi^2$ & $C_{15}^\prime m_\pi^4$  & $d_9^\prime m_\pi^2$ \\ 
      \hline
        NVIa$^\star$  & $-0.050(2)$ & $0.49(7)$ & $0.094(4)$ & $0.012(1)$ & $0.023(7)$\\
      \hline
        NVIb$^\star$  & $-0.055(3)$ & $0.09(5)$ & $0.073(3)$ & $0.025(2)$ & $ 0.030(6)$  \\
      \hline
        NVIIa$^\star$ & $-0.066(2)$ & $0.01(7)$ & $0.069(4)$ & $0.011(1)$ & $0.019(7)$ \\
      \hline
        NVIIb$^\star$ & $-0.049(2)$ & $0.09(4)$ & $0.048(3)$ & $0.017(1)$ & $0.018(3)$ \\
      \hline\hline
    \end{tabular}
    \caption{\label{tab:fitNV} Values of the
      LECs corresponding to the NV2+3 Hamiltonians Ia$^\star$, Ib$^\star$, IIa$^\star$, and IIb$^\star$ obtained in Ref.~\cite{Gnech:2022vwr} from fits to the deuteron and the trinucleon systems' magnetic moments, as well as deuteron photo-disintegration at backward angles data.}
  \end{center}
\end{table}

\section{Evaluation of the magnetic multipoles}
\label{sec:app1}
Using Eq.~\eqref{eq:mej} we can evaluate the value of the reduced matrix elements for the magnetic multipole operators. To do that we need a number of linear equations equal to the number of the multipoles allowed by the transition. 

As stated in Ref.~\cite{Carlson2015} for VMC and GFMC, in order to obtain the linear equations we need for the extraction of the reduced matrix elements, 
is more computationally efficient to change the direction of $\hat{\bm{q}}$ rather than repeating the calculation of the matrix elements for different projection of the total angular momentum. 
Therefore, if we have $n$ possible multipoles, we select $n$ values for the angle $\theta$ (that is the angle between the $\hat z$ axis and the direction of $\hat{\bm{q}}$) obtaining $n$ linear equations to invert. The choice of $\theta$ is arbitrary: we choose the ones that simplify as much as possible the solution of the linear equations. Below we report the solution of the linear equations for the cases studied in this work. To simplify the notation we indicate the matrix element computed using Monte Carlo as
\beq
X_\theta^J=\left\langle J\, J \left|{j}_{\gamma,y}(q(\theta))\right| J\,J  \right\rangle\,,
\eeq
and by
\beq
M_L^J=\left\langle J ||M_L|| J  \right\rangle\,,
\eeq
the reduced matrix elements of the magnetic multipole $L$ among states $J$.

For nuclei with $J=1/2$ and $J=1$ only the multipole with $L=1$ is allowed. In this case we need only one equation for which we chose $\theta=\pi/2$ ($\hat{\bm{q}}\parallel\hat x$).
The resultant reduced matrix element for $J=1/2$ and $J=1$ is 
\beq
  M_1^{1/2,1}=\frac{1}{\sqrt{\pi}}X_{\pi/2}^{1/2,1}\,.
\eeq

The situation for the $J=\frac{3}{2}$ case is more involved since both the $M_1$ and $M_3$
multipoles contributes to the magnetic f.f. In order to obtain the two magnetic multipoles
we have to compute two matrix elements changing the direction of $\hat{\bm{q}}$. For this case we select
$\theta=\pi/2$ and $\theta=\pi/4$.  By inverting
the linear equations we obtain the magnetic multipoles as function of the computed matrix elements, as
\begin{align}
  M_1^{3/2}&=\sqrt{\frac{2}{5\pi}}\left[X^{3/2}_{\pi/2}+\frac{2\sqrt{2}}{3}X^{3/2}_{\pi/4}\right]\,,\\
  M_3^{3/2}&=\frac{8}{\sqrt{15\pi}}\left[X^{3/2}_{\pi/2}-\sqrt{2}X^{3/2}_{\pi/4}\right]\,.
\end{align}

For the $J=2$ case we use the same angle as $J=3/2$. The final result reads
\begin{align}
  M_1^{2}&=\frac{3}{2\sqrt{5\pi}}\left[X^{2}_{\pi/2}+\frac{2\sqrt{2}}{3}X^{2}_{\pi/4}\right]\,,\\
  M_3^{2}&=\frac{4\sqrt{2}}{\sqrt{15\pi}}\left[X^{2}_{\pi/2}-\sqrt{2}X^{2}_{\pi/4}\right]\,.
\end{align}

For $J=3$ three different multipoles with $L=1,~3$, and 5 can be excited. Therefore, we need to have three independent linear equations  that 
we obtain selecting $\theta=\pi/2$, $\theta=\pi/4$ and $\cos \theta=2\sqrt{2}/3$. The solution of the linear equations reads
\begin{align}
  M_1^{3}&=\sqrt{\frac{1}{7\pi}}\left[\frac{53}{280\sqrt{2}}X^3_{\pi/2}+\frac{232}{735}X^3_{\pi/4}-\frac{81}{1960\sqrt{2}}X^3_{1/3}\right]\,,\\
  M_3^{3}&=\frac{1}{\sqrt{\pi}}\left[\frac{13\sqrt{2}}{20}X^3_{\pi/2}-\frac{32}{35}X^3_{\pi/4}-\frac{81\sqrt{2}}{140}X^3_{1/3}\right]\,,\\
    M_5^{3}&=6\sqrt{\frac{35}{\pi}}\left[\frac{\sqrt{2}}{35}X^3_{\pi/2}-\frac{32}{245}X^3_{\pi/4}+\frac{27\sqrt{2}}{245}X^3_{1/3}\right]\,,
\end{align}
where $X^3_{1/3}=\bra{3\,3}{j}_{\gamma,y}(q(\cos\theta=1/3))\ket{3\,3}$.

\section{Note on the magnetic form factors data}
The experimental data of the magnetic form factors shown in this work have been obtained directly the original references.
The definitions used in some of the experimental works for the magnetic f.f. are different compared to the one used in this work, therefore we renormalize them with an appropriate factor given by the ratio among the experimental values $F^{2*}_M(q)$ in the paper and the one defined in Eq.~\eqref{eq:fm}. For example, in the cases of  Refs.~\cite{Vanpraet1965,Rand1966,Vanniftrik1971}, this ratio is given by
\beq\label{eq:ratio_ex}
\frac{F^2_M(q)}{F^{2*}_M(q)}=\frac{1}{2\pi}
   \mu^2\left(\frac{q}{2M}\right)^2\frac{J+1}{3J}\,,
   \eeq
   where $\mu$ is the magnetic moment and $J$ the total angular momentum of the nucleus.
   
Note that Refs.~\cite{Goldemberg1963, Peterson1962,Lapikas1975,Goldemberg1965} report only the elastic scattering cross section data from which we obtain the values of the magnetic f.f. with the following formula
\begin{equation}\label{eq:xsec-ff}
  F_M^{2}(q)=\left(\frac{d\sigma}{d\Omega}\right)_{Exp}
  \left[4\pi \left(\frac{\alpha}{2E_0}\right)^2
    \right]^{-1}\,,
\end{equation}
where $\alpha$ is the fine structure constant, and $E_0$ the energy of the electron beam.
Note that the data of Refs.~\cite{Goldemberg1963, Peterson1962,Goldemberg1965} tend to overestimate the other experimental data sets. Indeed we excluded them in the plots of ${}^{10}$B magnetic f.f., following what was done in Ref.~\cite{Donnelly1984}.

The data for ${}^3$H and ${}^3$He are from a global fit performed in Ref.~\cite{Amroun:1994qj} of the electron scattering cross sections at different angles applying the Rosenbluth separation, which was then reused in~\cite{Sick2001} where the data are taken from.
For Refs.~\cite{Vanniftrik1971,Lapikas1978}, no numerical values were available, and therefore we digitized directly the plots on the paper using {\it Web Plot Digitizer}~\cite{Rohatgi2022}.
In Table~\ref{tab:data}, we report for each nucleus and experimental data set the reference, the normalization compared to our definition of the magnetic f.f. or the formula used to extract from the cross section respect to the original reference, and if the data have been digitized (D) or taken as numeric (N). The digitized data sets are available as Supplemental Material to this paper.

\begin{table}
\begin{center}
\begin{tabular}{llcc}
  \hline
  \hline
Nucleus & Reference & Data type & ratio/method \\   
  \hline
   ${}^3$H & Sick 2001~\cite{Sick2001} & N & 1 \\
   & & & \\
  ${}^3$He & Sick 2001~\cite{Sick2001} & N & 1 \\
   & & & \\
  ${}^6$Li & Peterson 1962~\cite{Peterson1962} & N & Eq.~\eqref{eq:xsec-ff} \\
           & Goldemberg 1963~\cite{Goldemberg1963} & N & Eq.~\eqref{eq:xsec-ff} \\
           & Rand 1966~\cite{Rand1966} & N & Eq.~\eqref{eq:ratio_ex}\\
           & Lapikas 1978~\cite{Lapikas1978} & D & $1/4\pi$ \\
           & Bergstrom 1982~\cite{Bergstrom1982} & N & $Z^2/4\pi$\\
            & & & \\
    ${}^7$Li & Peterson 1962~\cite{Peterson1962} & N & Eq.~\eqref{eq:xsec-ff} \\
    & Goldemberg 1963~\cite{Goldemberg1963} & N & Eq.~\eqref{eq:xsec-ff} \\
    &Van Niftrik 1971~\cite{Vanniftrik1971} & D & Eq.~\eqref{eq:ratio_ex}\\
    &Lichtenstadt 1983~\cite{Lichtenstadt1983} & N & $Z^2/4\pi$\\
     & & & \\
    ${}^9$Be & Goldemberg 1963~\cite{Goldemberg1963} & N & Eq.~\eqref{eq:xsec-ff}\\
    & Vanpraet 1965~\cite{Vanpraet1965} & N & Eq.~\eqref{eq:ratio_ex} \\
    & Rand 1966~\cite{Rand1966} & N & Eq.~\eqref{eq:ratio_ex}\\
    & Lapikas 1975~\cite{Lapikas1975} & N& Eq.~\eqref{eq:xsec-ff} \\
     & & & \\
     ${}^{10}$B & Goldemberg 1963~\cite{Goldemberg1963} & N & Eq.~\eqref{eq:xsec-ff} \\
     & Goldemberg 1965~\cite{Goldemberg1965} & N & Eq.~\eqref{eq:xsec-ff}\\
     &Vanpraet 1965~\cite{Vanpraet1965} & N  & Eq.~\eqref{eq:ratio_ex}\\
     & Rand 1966~\cite{Rand1966} & N & Eq.~\eqref{eq:ratio_ex}\\
     & Lapikas 1978~\cite{Lapikas1978} & D & $1/4\pi$ \\
  \hline
  \hline
\end{tabular}
\end{center}
\caption{\label{tab:data}
  Summary table of experimental data on elastic magnetic electron scattering reported in this work.  For each nucleus and experimental data set  the reference, the normalization or method used to obtain the form factor from to the original reference, and if the data have been digitized (D) or taken as numeric (N) is shown. We indicate the charge of the nucleus with $Z$. }
\end{table}

\bibliography{biblio,bib_magn}

\end{document}